\documentclass[pdflatex,sn-mathphys-num]{sn-jnl}

\usepackage{fix-cm}
\usepackage{graphicx}%
\usepackage{multirow}%
\usepackage{amsmath,amssymb,amsfonts}%
\usepackage{amsthm}%
\usepackage[mathscr]{eucal}
\usepackage[title]{appendix}%
\usepackage{xcolor}%
\usepackage{textcomp}%
\usepackage{manyfoot}%
\usepackage{booktabs}%
\usepackage{algorithm}%
\usepackage{algorithmicx}%
\usepackage{algpseudocode}%
\usepackage{listings}%
\usepackage{footmisc}

\usepackage{lineno}
\usepackage{hyperref}
\usepackage{braket}
\def\Ketbra#1#2{\left|#1\middle\rangle\middle\langle#2\right|}

\newcommand{\HI}{\text{H\fontsize{8}{0}\selectfont \,I }}

\begin{document}

\title[Subradiance and dark matter]{Quantum coherence and the invisible Universe: Subradiance as a dark matter mechanism}


\author*[1]{\sur{Martin} \fnm{Houde}}\email{mhoude2@uwo.ca}
\author[2]{\sur{Fereshteh} \fnm{Rajabi}}\email{rajabf1@mcmaster.ca}
\author[1]{\sur{Lamies} \fnm{Sati}}
\author[1,2]{\sur{Vahid} \fnm{Anari}}


\affil*[1]{\orgdiv{Department of Physics and Astronomy}, \orgname{The University of Western Ontario}, \orgaddress{\street{1151 Richmond Street}, \city{London}, \postcode{N6A 3K7}, \state{Ontario}, \country{Canada}}}

\affil[2]{\orgdiv{Department of Physics and Astronomy}, \orgname{McMaster University}, \orgaddress{\street{1280 Main Street West}, \city{Hamilton}, \postcode{L8S 4L8}, \state{Ontario}, \country{Canada}}}



\abstract{The origin of dark matter in galactic halos, one of the deepest unsolved problems in astrophysics, may find an unexpected contribution from the quantum mechanics of ordinary atomic hydrogen. We show that quantum entanglement and coherence among hydrogen atoms in a gas at thermal equilibrium can naturally lead to subradiance, a cooperative suppression of radiation that renders the gas simultaneously dark in emission, transparent to incident radiation, and effectively collision-less. These three properties, precisely those associated with dark matter, emerge from a single underlying physical mechanism: the entangled structure of Dicke states in the gas. Applying this framework to the 21 cm line of atomic hydrogen in galactic dark matter halos, we find that conditions there place the gas deep in the asymptotic subradiance regime, where the strongly suppressed spontaneous and stimulated intensities cancel exactly. The cold neutral cores of high-velocity clouds, with their observed temperatures near 100 K and inferred dark-to-visible mass ratios of $\sim$100:1, are consistent with this picture. Our results suggest that a significant fraction of the non-luminous matter pervading galactic halos may be familiar atomic hydrogen whose quantum cooperative behavior hides it from view; a solution that may have been hiding in plain sight.}




\maketitle

\section{Introduction}\label{sec1}

It has been known since Dicke's original treatment of the spontaneous emission of coherent radiation \cite{Dicke1954} that the interaction between atoms composing a gas through their common electromagnetic field can lead to the formation of entangled quantum mechanical states, with significant implications for the spontaneous emission rates of photons emanating from the gas. As Dicke pointed out, atoms should not be considered as independent entities; instead, the entire gas must be treated as a single quantum mechanical system. While the material covered in his paper was broad, touching on several phenomena involving coherence and entanglement, much of the early work it inspired focused on superradiance.

Superradiance is a fundamental radiation process \cite{Feld1980} in which the cooperative behavior of the atoms composing the gas leads to powerful bursts of coherent spontaneous emission. Indeed, superradiance can be thought of as “collective spontaneous emission,” where, instead of spontaneously emitting photons individually, the atoms cooperate in a coherent cascade of photon emission. Although the first experimental verification of superradiance occurred almost 20 years after Dicke's work \cite{Skribanowitz1973}, astronomers remained largely unaware of this phenomenon and the intense research being conducted in the quantum optics community. This remained the case despite the earlier discovery of astronomical masers in the interstellar medium \cite{Weaver1965} and the complementarity between the maser action and superradiance \cite{Feld1980,Rajabi2020}. It is only recently that superradiance has been studied in an astronomical context and used to explain intense flares from maser-hosting regions in diverse environments \cite{Rajabi2016A,Rajabi2016B,Rajabi2017,Houde2018a,Houde2019,Rajabi2019,Rajabi2020,Rajabi2023,Houde2024,Rashidi2025,Rashidi2026}. Beyond improving our understanding and modeling of these systems, these studies have importantly established the presence of the type of coherence proposed by Dicke in astronomical media.

Dicke's 1954 paper \cite{Dicke1954} also explicitly discussed another coherent radiation process, which later became known as subradiance. In contrast to superradiance, subradiance describes the suppression of photon emission compared to the typical spontaneous emission process. Dicke analyzed the trapping of energy in slow and dark quantum states and its consequences for the radiation intensity emanating from a gas. Subradiance was experimentally verified later than superradiance \cite{Pavolini1985} and has only recently been observed in large clouds of cold atoms \cite{Guerin2016,Das2020_sub,Cipris2021}. The latter experimental results are particularly relevant to us as we explore the possible existence of subradiant systems in astrophysical contexts.

More specifically, we ask the following question: Given that superradiance is known to occur in diverse astronomical environments, should we not also expect subradiance to be realized? We aim to answer this by exploring a possible link between subradiance and the presence of dark matter in galactic halos. For this, we consider the radiative properties associated with the atomic hydrogen 21~cm line in High Velocity \HI Clouds (HVCs) that permeate these environments. More specifically, we focus on the cold cores in HVCs known to host temperatures of approximately 100~K while exhibiting a dark matter-to-\HI mass ratio of $\sim$ 100:1 \cite{Wolfire1995,Bruns2000,Thilker2004}. These structures provide fertile environments to study a potential connection between subradiance and the dark matter phenomenon -- one of the deepest mysteries in astrophysics since its manifestation in galactic rotation curves was discovered \cite{Rubin1970}.


\section{Results}\label{sec:results}
\subsection{Entangled quantum states and coherent behavior}\label{sec:entangled state}

In this section we revisit parts of Dicke's work and focus on the trapping of energy within an atomic ensemble. We first focus on a two-atom system at resonance with Atom 1 at $-z_0/2$ and Atom 2 at $z_0/2$ on the $z$-axis. Given that single-atom transitions take place between the lower $\Ket{a}$ and upper $\Ket{b}$ levels of energies $-\hbar\omega_0/2$ and $\hbar\omega_0/2$, respectively, appropriate internal states for the two-atom problem consist of \cite{Dicke1954,Dicke1964} (see Appendices \ref{app:Hamiltonian} and \ref{app:cross_section})
\begin{align}
    \Ket{1,1} &= \Ket{bb} \label{eq:|1,1>}\\
    \Ket{1,0}_\theta &= \frac{1}{\sqrt 2}\left(e^{i\frac{1}{2}k z_0 \cos\theta}\Ket{ab} + e^{-i\frac{1}{2}k z_0 \cos\theta}\Ket{ba}\right) \label{eq:|1,0>}\\
    \Ket{1,-1} &= \Ket{aa} \label{eq:|1,-1>}\\
    \Ket{0,0}_\theta &= \frac{1}{\sqrt 2}\left(e^{i\frac{1}{2}k z_0 \cos\theta}\Ket{ab} - e^{-i\frac{1}{2}k z_0 \cos\theta}\Ket{ba}\right). \label{eq:|0,0>}
\end{align}

\noindent Following Dicke \cite{Dicke1954}, the kets $\Ket{r,m}$ are defined with the ``cooperative'' number $r$ and ``inversion'' number $m$, which are analogous to quantum numbers for a system of coupled spin-1/2 particles, with $r\ge 0$ and $\left|m\right|\le r$ ($\Ket{r,m}_{\theta}=\Ket{r,m}$ when $\cos\theta=0$ with $\theta$ the angle between the $\mathbf{k}$-vector and the $z$-axis). Note that $k=\left|\mathbf{k}\right|=2\pi/\lambda$ with $\lambda$ the radiation wavelength.

Given equations (\ref{eq:|1,1>})-(\ref{eq:|0,0>}) it is straightforward to calculate the spontaneous emission transition rates per unit solid angle between the different states as
\begin{align}
    \frac{d\gamma_{1,1\rightarrow 1,0_\theta}}{d\Omega} &= 2\frac{d\Gamma}{d\Omega} \label{eq:11->10} \\
    \frac{d\gamma_{1,1\rightarrow 0,0_\theta}}{d\Omega} &= 0 \label{eq:11->00} \\
    \frac{d\gamma_{1,0_\theta\rightarrow 1,-1}}{d\Omega^\prime} &= 2\cos^2\left[\frac{1}{2}k z_0\left(\cos\theta^\prime-\cos\theta\right)\right]\frac{d\Gamma}{d\Omega^\prime} \label{eq:10->1-1} \\
    \frac{d\gamma_{0,0_\theta\rightarrow 1,-1}}{d\Omega^\prime} &= 2\sin^2\left[\frac{1}{2}k z_0\left(\cos\theta^\prime-\cos\theta\right)\right]\frac{d\Gamma}{d\Omega^\prime} \label{eq:00->1-1} 
\end{align}
with $d\Gamma/d\Omega$ the spontaneous emission rate per unit solid angle of a single independent atom, and $\theta$ and $\theta^\prime$ denoting the orientations of the first and second emitted photons, respectively, relative to the $z$-axis, as well as $\Omega$ and $\Omega'$ the corresponding solid angles. 

The signatures of entanglement and coherence are displayed in equations (\ref{eq:10->1-1})-(\ref{eq:00->1-1}). That is, the enhancement of the transitions rates by a factor of `2' is due to the entangled nature of the states given in equations (\ref{eq:|1,0>}) and (\ref{eq:|0,0>}), while the disappearance of the argument in the cosine and sine functions (i.e., when $\cos\theta^\prime = \cos\theta$) stems from coherence between the two emitted photons. We therefore find the existence of an angular correlation between successively emitted photons. More precisely, if the system starts in the $\Ket{1,1}$ state, then after the emission of the first photon it will for sure end up in the symmetric $\Ket{1,0}_\theta$ intermediate state. From there, we see that the second photon has a higher probability of being emitted in a direction where $\cos\theta^\prime = \cos\theta$ to reach the $\Ket{1,-1}$ ground state \cite{Dicke1964}.

We will soon be interested in situations where the system is in thermal equilibrium where all the states for a given $m$ value are initially equally populated. We thus focus on the transition rates between the $m=0$ and $m=-1$ states given in equations (\ref{eq:10->1-1})-(\ref{eq:00->1-1}). We can obtain the radiation intensity by, among other things, integrating these transition rates over the solid angle. Setting $\cos\theta=0$, for simplicity, and initial conditions where the states $\Ket{1,0}_\theta$ and $\Ket{0,0}_\theta$ have equal but uncorrelated probabilities of occupation (of $1/2$), we can study the intensity in two opposite limits (see Appendix \ref{app:two-atom}).  

For the infinite size sample $k z_0\rightarrow\infty$ we find 
\begin{equation}
    I_\infty\left(t\right) = \hbar\omega_0\Gamma e^{-\Gamma t}, \label{eq:I_inf}
\end{equation}
\noindent which is exactly the intensity from a single independent atom, with $\Gamma$ the single-atom free-space spontaneous emission rate (see equation \ref{eq:Gamma}). 

On the other hand, in the idealized small sample limit $kz_0=0$ the intensity becomes
\begin{equation}
    I_\mathrm{ss}\left(t\right) = \frac{1}{2}\hbar\omega_0\left(2\Gamma\right) e^{-2\Gamma t}. \label{eq:I_ss}
\end{equation}
The two-atom system will, on average, emit a photon half of the time but at twice the single-atom rate. Another way to look at this is by considering the energy trapping efficiency defined as
\begin{align}
    \eta & = 1-\frac{\int_0^\infty I_\mathrm{ss}\left(t\right)dt}{\int_0^\infty I_\infty\left(t\right)dt} \nonumber \\
    & = 0.5. \label{eq:eta}
\end{align}
\noindent Again, we find that half of the internal energy initially stored in the system remains within it in the steady-state. 

The small sample behavior is readily understood from equations (\ref{eq:10->1-1})-(\ref{eq:00->1-1}) since in this case coherence happens for all radiation modes, resulting in the doubling and cancellation of the corresponding transition rates. The latter implies that $\Ket{0,0}_\theta$ is a \textit{dark state} and that its initial occupation probability is preserved at all times. This is where energy trapping and subradiance originate. On the other hand, the behavior for $k z_0\rightarrow\infty$ is a global characteristic in the sense that it can only be discerned in the solid-angle-integrated transition rate. Even at very large separations some radiation modes will be coherent, but when considering all modes the two atoms behave as if they were independent  (see equation \ref{eq:g00->1-1_app}).


We also note that \textit{superabsorption}, i.e., the converse of superradiance, takes place when the two-atom system is initially in the $\Ket{1,-1}$ ground state. When subjected to an incident one-photon field the transition rates of equations (\ref{eq:11->10})-(\ref{eq:00->1-1}) are reversed \cite{Yang2021}. Here again, there is angular correlation, but this time between successively absorbed photons (see Appendix \ref{app:two-atom}). Likewise, \textit{subabsorption} will result when the system is initially in the $\Ket{0,0}_\theta$ state.

\begin{figure}
    \centering 
    \includegraphics[width=0.85\columnwidth]{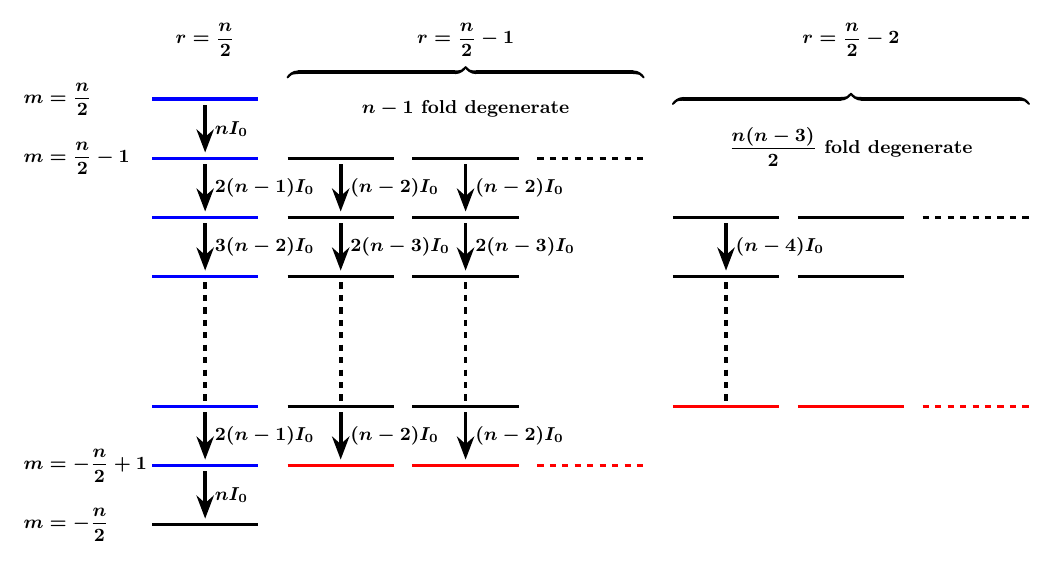}
    \caption{The coherent radiation rates between Dicke states in the small sample (i.e., when $kz_0\rightarrow 0$) or coherent limits. The number of atom contained in the radiating gas is $n$, while the states within the different coherent cascades are identified through the cooperative number $0\leq r\leq n/2$ and the inversion number $-r\leq m\leq r$. The radiative intensity (or rate) of a single atom is denoted by $I_0$.  The symmetric and dark edge states (excluding the ground state) are shown in blue and red, respectively (see Sec.~\ref{sec:master equation}). Adapted from Dicke's original 1954 paper \cite{Dicke1954}.}
    \label{fig:Dicke-table}
\end{figure}

The calculations can be extended to a larger number of atoms. The general case for an arbitrary number of atoms $n$ is shown in Figure \ref{fig:Dicke-table} along with the expected transition rates in the small sample (i.e., $kz_0\rightarrow 0$) or coherent limits (adapted from Dicke's original paper \cite{Dicke1954}). It is generally observed that, starting from a given $m$ value, transitions are allowed only between states sharing the same cooperation number $r$, while rates systematically decrease as one moves rightward to lower $r$ values. 

It is shown in Appendix \ref{sec:n-atom} that for $n\gg 1$ the energy trapping efficiency is $\eta\approx 1-\sqrt{2/n}$ when starting from $m=0$. It is therefore clear that the combination of coherence between the collective states of a large number of atoms can potentially lead to a significant amount of energy trapping within the gas. But one of our goals is to assess the level of energy trapping within the gas for finite and more realistic atomic separations in astrophysical environments. It is already obvious through our study of the two-atom system that the general case where $kz_0\neq 0$ will lead to a non-zero transition rate from the $\Ket{0,0}_\theta$ `dark' state (see equations \ref{eq:00->1-1} and \ref{eq:g00->1-1_app}). This has for implication that all the energy initially stored in the system will eventually leak away. That is, $\eta=0$ in the steady-state and no energy trapping is realized in the long run.

\subsection{Equilibrium conditions}\label{sec:master equation}

To further our study of trapping and leaking of internal energy we move to the density matrix framework, using a master equation that is a generalization of one often used within the context of superradiance \cite{Lehmberg1970,Gross1982} (see Appendix \ref{app:master equation}). Instead of modeling the environment with the radiation vacuum state, this generalized master equation allows for the existence of one non-empty radiation state $\ket{n_{k^\prime}}$ containing $n_{k^\prime}$ photons of mode $\mathbf{k}^\prime=\left(\omega_0/c\right)\boldsymbol{\epsilon}_{k^\prime}$ (i.e., $\left|\mathbf{k}^\prime\right|=k$). Later on, this photon state will be used to model a  non-coherent incident radiation field.

\begin{figure}
    \centering 
    \includegraphics[width=0.75\columnwidth]{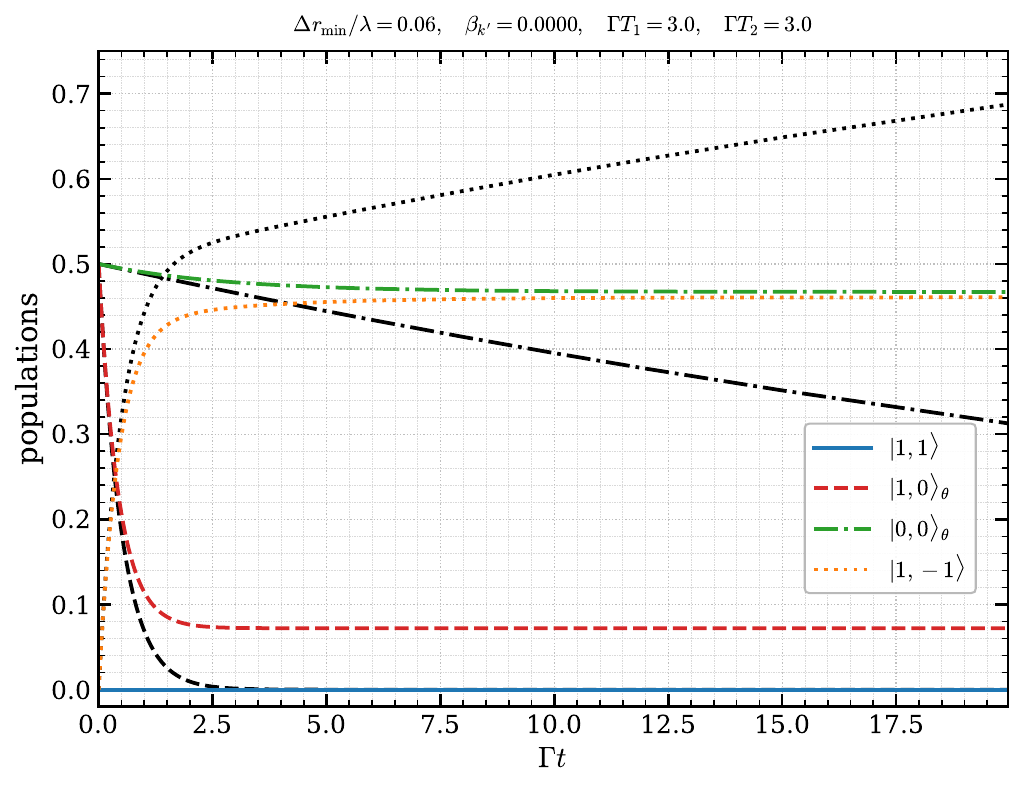}
    \caption{Solution of the two-atom system under the initial conditions where the populations for the $\Ket{1,0}_\theta$ and $\Ket{0,0}_\theta$ $m=0$ states are set to $1/2$ with all other populations and coherences to zero. The separation between the atoms was set to $\Delta r=0.06\lambda$. The time evolution of the relevant populations are traced using the legend for the corresponding states. \textit{Black curves:} no equilibrium conditions or relaxation/dephasing terms are applied and the populations corresponding to $m=0$ will eventually all relax to zero. No energy trapping takes place in the steady-state (i.e., when $\Gamma t\rightarrow\infty$). \textit{Colored curves:} Pumps, enforcing equilibrium populations levels of 1/2 on the $m=0$ states, acting on the $T_1$ time-scale, as well as relaxation and dephasing terms $\Gamma T_1=\Gamma T_2=3$, are applied. Leakage is prevented and energy is trapped in the steady-state.} 
    \label{fig:two-atom}
\end{figure}

For our two-atom problem, the evolution of the density matrix $\hat{\rho}$ for the atomic system will be defined using the states given in equations (\ref{eq:|1,1>})-(\ref{eq:|0,0>}) (we set $\cos\theta=0$ for simplification). The solution of the master equation for the two-atom problem discussed in Sec.~\ref{sec:entangled state} (and in Appendix \ref{app:master equation}) is presented in Figure \ref{fig:two-atom} for an atomic separation $\Delta r=0.06\lambda$, which corresponds to a density of $\approx 0.5\;\mathrm{cm}^{-3}$ at $\lambda=21$~cm. The black curves trace the time evolution of the corresponding populations when no incident radiation field or relaxation/dephasing terms are applied (see below and Appendix \ref{app:master equation}). For initial conditions, we set the populations of the $m=0$ states to $1/2$, and all other populations and coherences to zero (see below). As can be inferred from the figure, no energy is trapped in the system as the populations corresponding to the $\Ket{1,0}_\theta$ and $\Ket{0,0}_\theta$ states go to zero in the steady-state (i.e., as $\Gamma t\rightarrow\infty$). The first (fast) state does so quickly (on a time-scale of $\Gamma t\sim 1/2$) while the second (dark) state decays on a much longer time-scale, but will still eventually go down to zero. 

As we will now show, the leakage of internal energy is stopped in the steady-state by enforcing equilibrium conditions for the populations. At thermal equilibrium, the initial conditions result from the environment the atoms find themselves in, where not all transitions are due to radiative processes. Collisions, for example, can contribute to establish conditions where all states sharing the same energy level will initially be populated equally. Furthermore, and as was shown by Dicke \cite{Dicke1954}, for systems containing a large number of atoms $n$ with $\hbar\omega_0\ll k_\mathrm{B}T_\mathrm{kin}$ (as for the 21~cm line in astrophysical settings; $k_\mathrm{B}$ is the Boltzmann constant and $T_\mathrm{kin}$ the kinetic temperature of the gas) thermal equilibrium leads to non-zero occupation of states at $\overline{m}=-n\hbar\omega_0/\left(4k_\mathrm{B}T_\mathrm{kin}\right)$ with negligible deviations from that level. While the two-atom system evidently does not contain a large number of atoms, for the sake of our discussion we apply this broader result by initially setting the $\Ket{1,0}_\theta$ and $\Ket{0,0}_\theta$ states equally with, again, all other populations and coherences to zero, and driving the $m=0$ states with pumps enforcing equilibrium (since $\overline{m}\approx0$). In Appendix \ref{app:master equation}, we show how this is implemented through the introduction of a relaxation term acting on a time-scale $T_1$ and a constant pump term, which we then define as $\rho_{jj}^\mathrm{eq}/T_1$, with $\rho_{jj}^\mathrm{eq}$ the equilibrium condition for the corresponding population. 

In Figure \ref{fig:two-atom} we also show, using the colored curves and legend, how the two-atom system responds when such pumps, enforcing equilibrium population levels of 1/2 on the $m=0$ states, are applied with $\Gamma T_1=\Gamma T_2=3$ for the pump/relaxation and dephasing time-scales, respectively\footnote{The time-scales $T_1$ and $T_2$ are given fiducial values in this section, but are equated to the collision time-scale at the kinetic temperature of the gas when applied to galactic halos later on in Sec.~\ref{sec:global entanglement}. Likewise, the equilibrium population levels $\rho^\mathrm{eq}_{jj}$ correspond to thermal equilibrium according to the Boltzmann formula.}. Since internal energy trapping is due to non-zero steady-state values in the corresponding populations, we find that the presence of equilibrium conditions effectively causes the trapping of energy to take place. More precisely, we find a $\simeq 0.46$ dark state population for an energy trapping efficiency $\eta\simeq 0.39$ in the steady-state\footnote{The energy trapping efficiency can also be defined in the steady-state by modifying equation (\ref{eq:eta}) to $\eta=1-I_\mathrm{sys}\left(t\rightarrow\infty\right)/I_\infty\left(t\rightarrow\infty\right)$, with $I_\mathrm{sys}$ the intensity of the system under consideration.}.  

\begin{figure}
    \centering 
    \includegraphics[width=0.75\columnwidth]{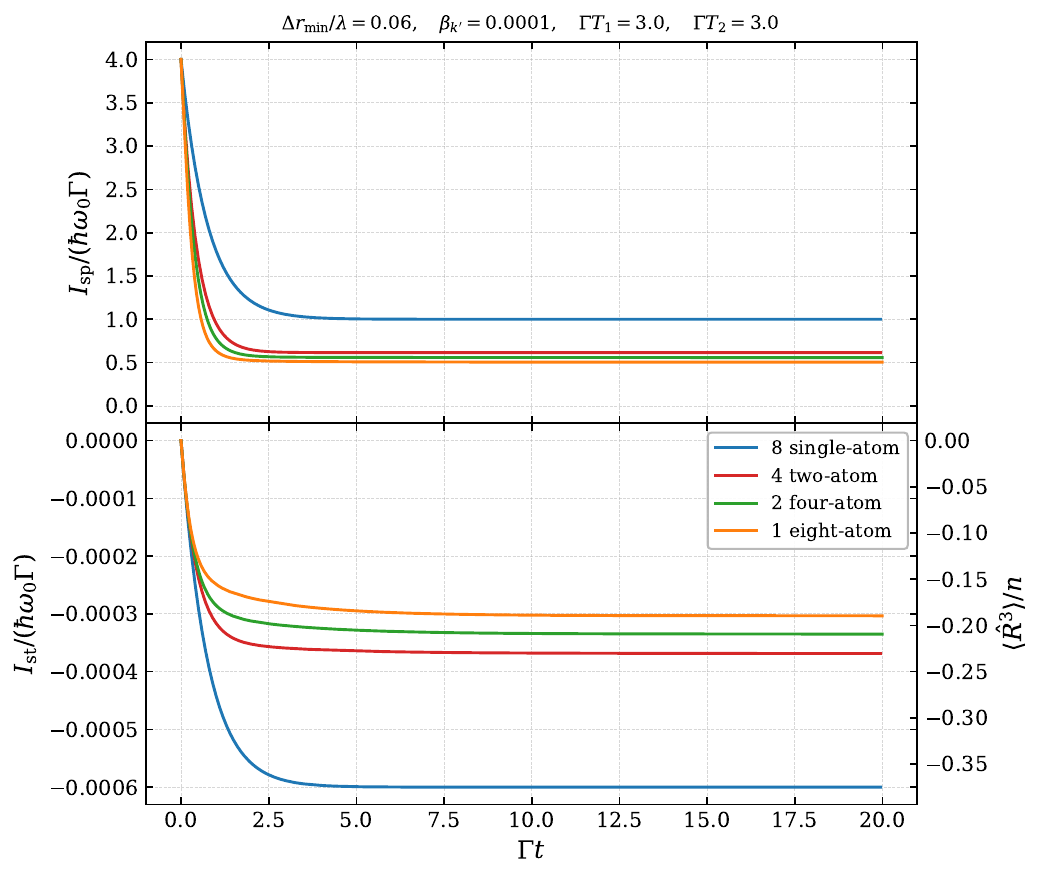}
    \caption{Spontaneous (top) and stimulated (bottom) intensities, normalized to $\hbar\omega_0\Gamma$, emanating from single-, two-, four- and eight-atom systems when $\Gamma T_1 = \Gamma T_2 = 3$ and a non-coherent incident field with $\beta_{k'}/\Gamma=10^{-4}$ is present. The mean excitation level per atom, $\left<\hat{R}^3\right>/n$, is given using the vertical axis on the right of the bottom graph. Subradiance and energy trapping are apparent from the narrower profiles of the transient intensities for the multi-atom systems over the single-atom case. As well, the spontaneous intensity decreases and the stimulated intensity becomes less negative with an increasing number of atoms in the steady-state (i.e., the system becomes darker in emission and more transparent to the incident radiation). The corresponding energy trapping efficiencies relative to a single-atom system are 0.39, 0.44 and 0.49 with increasing numbers of atoms.} 
    \label{fig:intensities}
\end{figure}

Such analysis can be extended to cases involving a larger number of atoms. Figure \ref{fig:intensities} shows the spontaneous and stimulated intensities (normalized to $\hbar\omega_0\Gamma$; see equations \ref{eq:Isp-coop}-\ref{eq:Ist-R3}), as well as the mean excitation level per atom (using the vertical axis on the right of the bottom graph), for the single-, two-, four- and eight-atom cases. The minimum separation between pairs of atoms is kept at $0.06\lambda$ with the four-atom system in a square configuration (in the $xz$-plane) and the eight-atom forming a cube (with its faces perpendicular to the $x$-, $y$- and $z$-axes), while we once again initially equally populated the $m=0$ states (all other populations and coherences are set to zero). The relaxation and dephasing time-scales are set to $\Gamma T_1 = \Gamma T_2 = 3$ and a non-coherent incident field propagating along the $y$-axis with $\beta_{k'}/\Gamma=10^{-4}$ is present. As explained in Appendix \ref{app:master equation}, the parameter 
\begin{align}
    \beta_{k^\prime} & = n_{k^\prime}\left(\frac{3\Gamma}{8\pi}\cdot\frac{\lambda}{L}\right) \label{eq:beta_k_main}
\end{align}
is proportional to the Einstein stimulated emission coefficient and approximately takes the adopted value for the radiation incident in a galactic dark matter halo at 21~cm for a typical spiral galaxy (see equation \ref{eq:beta_halo}). 

A few important features present in Figure \ref{fig:intensities} are worth discussing. First, while equilibrium conditions have enabled the trapping of energy in the long run, they also bring a non-zero steady-state intensity in all cases. This is not surprising since, as discussed above, equilibrium forces the establishing of finite steady-state populations. Second, we note that subradiance and energy trapping are apparent for the multi-atom systems in both the transient (from the narrower intensity profiles) and steady-state regimes. In the steady-state, the level of spontaneous intensity decreases with an increasing number of atoms, while the energy trapping increases from 0.39 to 0.44 and 0.49 when going from the two- to four- and eight-atom systems, respectively. On the other hand, the stimulated intensity and the mean excitation level per atom become less negative, implying a reduced level of stimulated absorption events. In other words, the systems become darker in emission and more transparent to the incident radiation as we increase the number of atoms.

\begin{figure}
    \centering 
    \includegraphics[width=0.75\columnwidth]{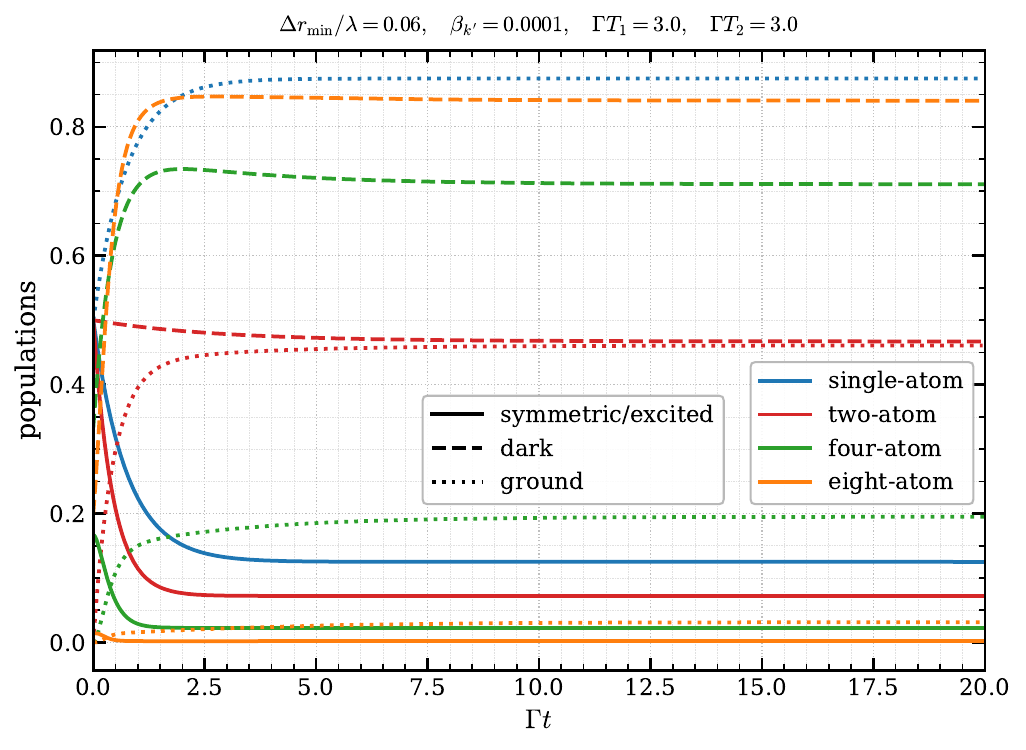}
    \caption{Population levels for the `symmetric' (`excited' for the single-atom), `ground' and `dark edge' (minimum $r$ at all $m\leq 0$ values except for the ground state; inexistent for the single-atom) states for the single-, the two-, four- and eight-atom systems of Figure \ref{fig:intensities}. The symmetric and ground states become increasingly depleted as the number of atom increases, while the dark edge states dominate in their combined population level due to their overwhelming degeneracy and low decay rates.} 
    \label{fig:pop}
\end{figure}

This behavior can be understood by observing the probability of occupations of different types of states in the steady-state regime. Accordingly, Figure \ref{fig:pop} shows the population levels for the `symmetric' (`excited' for the single-atom), `ground' and `dark edge' states for the single-, two-, four- and eight-atom systems of Figure \ref{fig:intensities}. The dark edge consists of all states for which $r$ is minimum at all $m\leq 0$ values (see Figure \ref{fig:Dicke-table}) except for the ground state; it is therefore inexistent for the single-atom. While the symmetric states (of maximum $r$ value; inhabiting the left column in Figure \ref{fig:Dicke-table}), which are responsible for the superradiance phenomenon, have the highest radiation rates at all $m$, the dark edge is entirely made up of the overwhelmingly degenerate dark states exhibiting the lowest decay rates. These characteristics lead to low population levels for symmetric states in the steady-state regime, and conversely for the dark edge (see equations \ref{eq:p11^ss}-\ref{eq:p22^ss}). These facts explain the trends seen in Figure \ref{fig:pop}: a systematic depletion of the symmetric and ground states, combined with a domination of dark edge states in their overall population level as the number of atoms increases. The multi-atom systems are thus bound to display a strong subradiance behavior. 

These results suggest that an extension to a gas containing a very large number of atoms may lead to an overwhelmingly suppressed level of emission and high transparency for the propagation of a non-coherent radiation field. These steady-state characteristics would result from the extensive depletion of the ground and symmetric states in favour of the strongly degenerate dark edge states.

\subsection{The continuum limit}\label{sec:MBE}

Extending the analysis of Sec.~\ref{sec:master equation} would quickly become computationally prohibitive for even a slight increase in the number of atoms. We thus move away from the discrete master equation framework and use a new set of equations which consists of the corresponding generalization to the continuum limit (see Appendix \ref{app:MBE}). 

We now study the propagation of an incident radiation field on a slab composed of an atomic gas, similar to what is often done in elementary radiative transfer analysis in astrophysics \cite{Rybicki1979}. The slab is of thickness $L$ along the $z$-axis and extends infinitely in perpendicular directions. This will allow us to compare the outcome of our analysis with known results. Within this context, the continuum equations determine the evolution of two quantities as a function of position $z$ and time $t$: the differential density $n'=n'_b -n'_a$, which corresponds to (half) of the population inversion with $n'_a$ and $n'_b$ (half) the densities of atoms in the lower and upper levels, respectively, and a function $\tilde{C}\left(z,z';t\right)$, which, when multiplied by the square of the dipole moment, stands for the two-point correlation function of the magnetization for a magnetic dipolar transition such as the 21~cm line (or the polarization for an electric dipole transition). Our equations include relaxation and dephasing time-scales $T_1$ and $T_2$, as well as a thermal equilibrium condition enforced on the time-scale $T_1$ through a corresponding differential density $n'_\mathrm{eq}=n'_{b,\mathrm{eq}} -n'_{a,\mathrm{eq}}<0$ set by the Boltzmann equation at the temperature of the gas $T_\mathrm{kin}$ (see equations \ref{eq:dn(z)}-\ref{eq:dC(z,z')}). This framework also allows us to independently trace the intensities due to spontaneous emission and stimulated processes (i.e., emission and absorption). As shown in Appendix \ref{app:MBE}, these intensities at $z=L$ are respectively given by (in units of energy per second per area)
\begin{align}
        I_\mathrm{sp}\left(L,t\right) &= \hbar\omega_0\frac{3\pi\Gamma}{2k^2}\int_0^Ldz\int_0^Ldz'\,\tilde{C}\left(z,z';t\right)\label{eq:Isp(L)_main}\\
    I_\mathrm{st}\left(L,t\right) &= 2\hbar\omega_0\beta_{k'}\int_0^Ldz\,n'\left(z,t\right).\label{eq:Ist(L)_main}
\end{align}

While the direction of propagation of the non-coherent incident field (i.e., along the $z$-axis) sets the symmetry axis of the slab, spontaneous emission into all other radiation modes is explicitly accounted for in the generalized master equation that serves as the basis for our continuum equations, as explained in Appendix \ref{app:MBE}. Specifically, the single-atom decay rate $\Gamma$
already encodes emission into the full solid angle of vacuum modes. The interaction between the incident field and the gas will elicit stimulated processes and favor coherence in subsequent transitions through angular correlation among photons, as discussed in Sec.~\ref{sec:entangled state}.

\subsubsection{Steady-state response in the linear regime}\label{sec:steady-state}

We are interested in determining the level of emission from the gas, as well as the results from its interaction with the incident field in the steady-state. Since we set the system at thermal equilibrium and are therefore expecting a relatively low intensity level, we assume the system to evolve in the linear regime where the population inversion is constant at $n'_\mathrm{eq}/\left(1+2\beta_{k'}T_1\right)<0$ (see Appendix \ref{app:linear}). We thus introduce the steady-state density
\begin{align}
    n'_0 = \frac{\left|n'_\mathrm{eq}\right|}{1+2\beta_{k'}T_1}>0,\label{ref:n'0_main} 
\end{align} 
which, in conjunction with equation (\ref{eq:Ist(L)_main}), yields
\begin{align}
    I_\mathrm{st}\left(L\right) = -2\hbar\omega_0\beta_{k'}n'_0 L\label{eq:Ist_steady_main}
\end{align}
for the stimulated intensity in the steady-state. The fact that $I_\mathrm{st}\left(L\right)<0$ is expected since absorption dominates over stimulated emission at thermal equilibrium. 

It is also shown in Appendix \ref{app:linear} that, under these conditions, 
\begin{align}
    \tilde{C}\left(z,z'\right) = \frac{2n'^2_0\beta_{k'}}{\alpha_{k'}}e^{-a\left(z+z'\right)}I_0\left(2a\sqrt{zz'}\right),\label{eq:C(z,z')_steady_main}  
\end{align}
where the \textit{subradiance coherence length-scale} is given by (see below) 
\begin{align}
    a^{-1}=\frac{2k^2\alpha_{k'}}{3\pi n'_0\Gamma}\label{eq:a}
\end{align}
with $\alpha_{k'}=\beta_{k'}+1/T_2$, and $I_0\left(x\right)$ the modified Bessel function of the first kind and order $0$ \cite{Abramowitz1972}. While equation (\ref{eq:C(z,z')_steady_main}) can be numerically integrated to yield the spontaneous intensity through equation (\ref{eq:Isp(L)_main}) (see Sec.~\ref{sec:global entanglement} below), the asymptotic limit $aL\gg 1$ can be analytically solved to yield
\begin{align}
        I_\mathrm{sp}\left(L\right) = 2\hbar\omega_0\beta_{k'}n'_0 L.\label{eq:Isp_asym_main}
\end{align}
We therefore find the remarkable outcome 
\begin{align}
    I_\mathrm{sp}\left(L\right) + I_\mathrm{st}\left(L\right) = 0\label{eq:Isp+Ist=0_main}
\end{align}
for $aL\gg 1$. In other words, when taken together the spontaneous and stimulated intensities are found to cancel each other exactly, resulting in a gas that appears completely dark in emission and perfectly transparent to the incident non-coherent radiation field. This result is consistent with the purported extension to a system containing a very large number of atoms discussed at the end of Sec.~\ref{sec:master equation}.

\section{Discussion -- subradiance in dark matter halos}\label{sec:global entanglement}

We now apply our analysis to galactic dark matter halos to determine whether the 21~cm line of atomic hydrogen could account for the known characteristics associated to dark matter. Since the presence of atomic hydrogen is clearly established observationally through the detection of 21~cm signals at distances $>100$~kpc deep into galactic halos\footnotemark{} \cite{Thilker2004,Das2020,Das2024}, the question to answer is whether the physical process discussed in this paper could hide most of the atomic hydrogen and account for the dark matter known to exist in these environments. 

\begin{figure}
    \centering 
    \includegraphics[width=0.75\columnwidth]{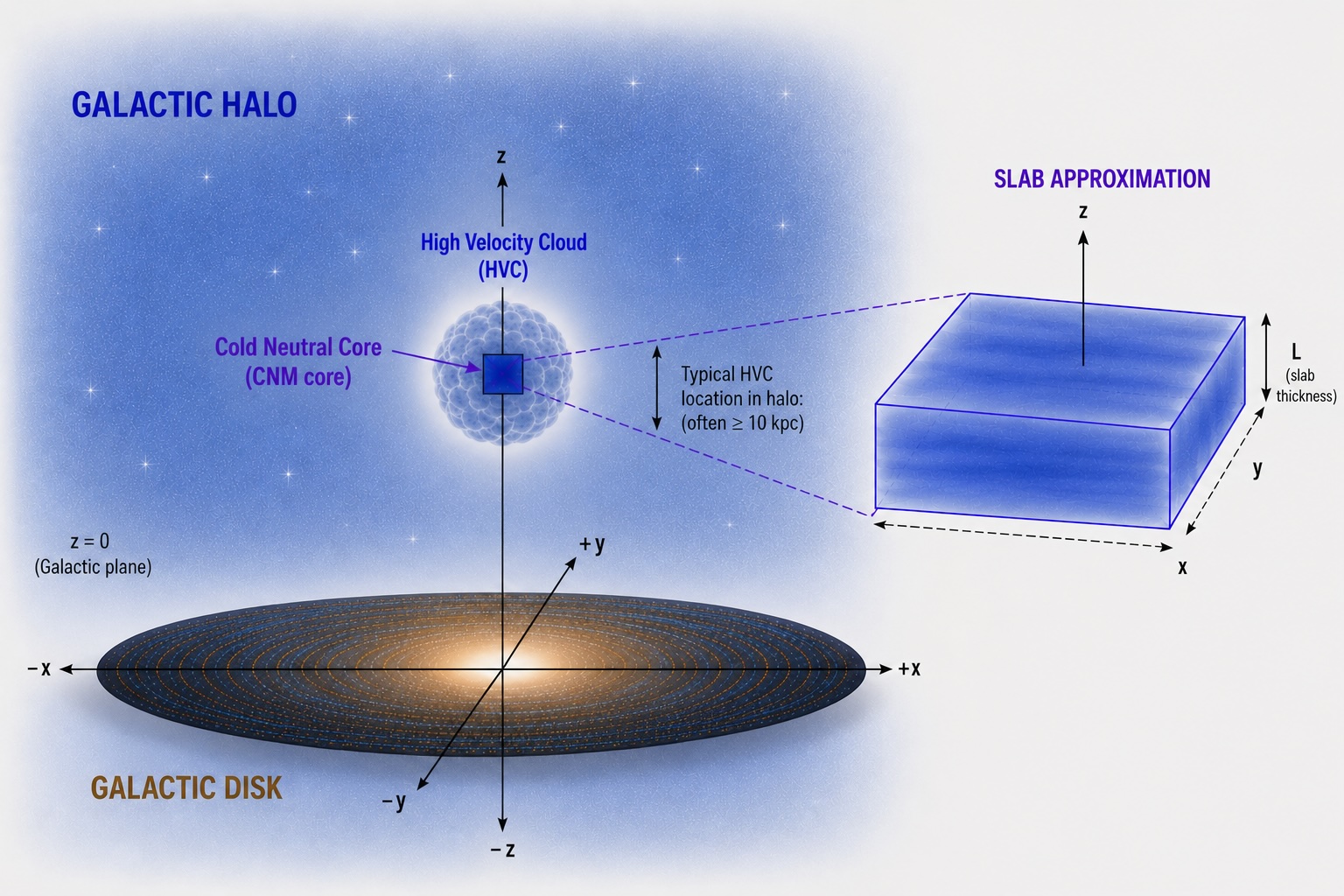}
    \caption{Schematic of the galactic environment considered in this work. An HVC is located in a galactic halo at a typical height of $\gtrsim 10$~kpc above the galactic plane ($z=0$). The CNM core embedded within the HVC provides the physical conditions required for subradiance: sufficiently high density, low kinetic temperature ($\sim 100$~K), and a directed radiation field from the galactic disk below. The inset shows the slab approximation adopted in our analysis, in which the CNM core is modeled as a plane-parallel slab of thickness $L$ oriented perpendicular to the $z$-axis, extending infinitely in the $x$- and $y$-directions. Dimensions are not to scale.} 
    \label{fig:slab}
\end{figure}

To answer this question we must set some defining parameters for our model. As was done in previous sections, we adopt an atomic hydrogen density $0.5\;\mathrm{cm}^{-3}$, which approximately corresponds to the mean mass density of dark matter measured in the vicinity of the Sun and in galaxy clusters ($\approx 0.5\;\mathrm{GeV\,cm^{-3}}$; see \cite{Cirelli2024}). This value is also consistent with measurements obtained for the cold neutral medium (CNM) of HVCs \cite{Marchal2021}. While high-resolution observations have established that HVCs possess a two-phase structure: CNM cores at kinetic temperatures near 100 K embedded within warm neutral medium (WNM) envelopes at  $\sim 8000$~K \cite{Wolfire1995,Bruns2000,Braun2000}, within our framework, the CNM cores provide the conditions and parameters chosen for our analysis. We therefore set a gas temperature $T_\mathrm{kin}=100$~K, which leads to $n'_0=8.5\times 10^{-5}\;\mathrm{cm}^{-3}$ and $T_1=T_2=1.6\times 10^9$~s when assuming that collisions set these time-scales. As discussed at the end of Appendix \ref{app:master equation}, we posit an incident 1.4~GHz flux density of $10^6$~Jy at a distance of 10~kpc in the halo\footnotemark[\value{footnote}]\footnotetext{$1\;\mathrm{pc}=3.09\times10^{18}$~cm and $1\;\mathrm{Jy}=10^{-26}\;\mathrm{W\,m^{-2}\,Hz^{-1}}$.}, a suitable location for HVCs. This implies $\beta_{k'}/\Gamma=5.5\times 10^{-5}$, $\alpha_{k'}\simeq 1/T_2=6.4\times 10^{-10}\;\mathrm{s}^{-1}$ and a length-scale $a^{-1}=5\times 10^7$~cm. Given the sheer size of dark matter halos and structures within them (for example, HVCs are known to have a characteristic size on the order of 0.5~kpc \cite{Thilker2004}), it follows that astrophysical regimes are deep into the asymptotic limit $aL\gg 1$. We show in Figure \ref{fig:slab} a sketch of the galactic environment we consider, where an HVC is located in a galactic halo above the galactic plane with an embedded CNM core modeled as a plane-parallel slab of thickness $L$. 

\begin{figure}
    \centering 
    \includegraphics[width=0.75\columnwidth]{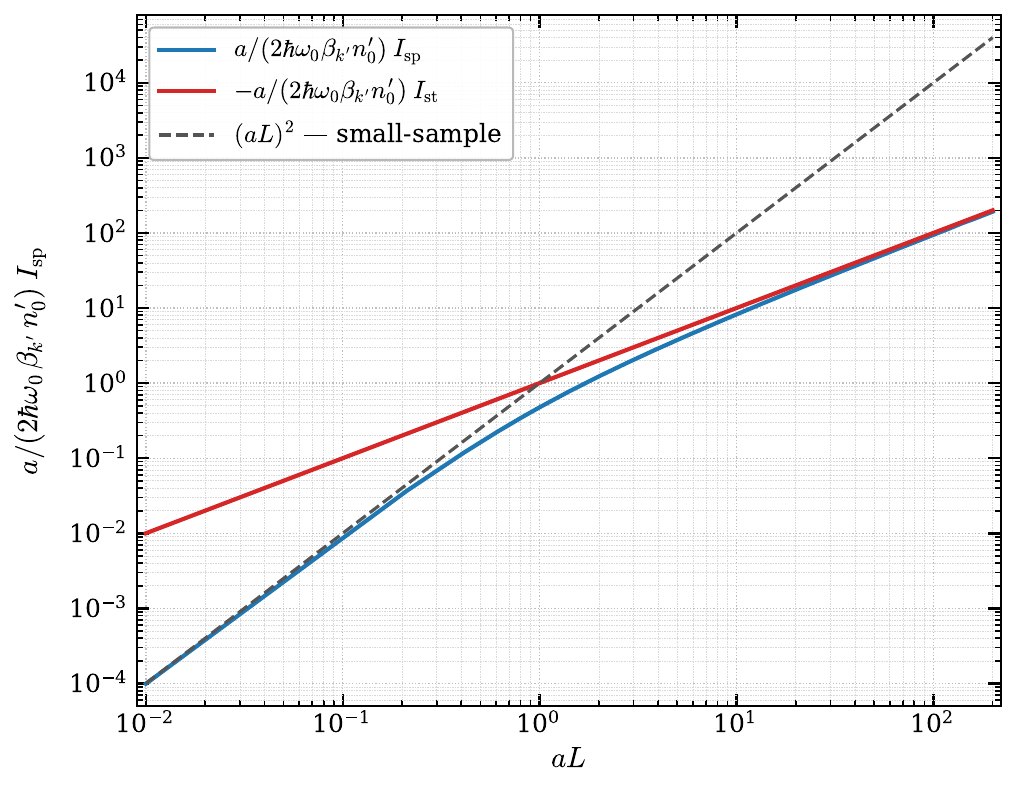}
    \caption{The steady-state intensity due to spontaneous emission $I_\mathrm{sp}$ (normalized to $2\hbar\omega_0\beta_{k'}n'_0/a$; blue curve) as a function of the normalized distance $aL$, as calculated with equations (\ref{eq:Isp(L)_main}) and (\ref{eq:C(z,z')_steady_main}). Also shown are $-I_\mathrm{st}$, the negative of the stimulated intensity (red line), and the small-sample limit of $I_\mathrm{sp}\propto \left(aL\right)^2$, when $aL\ll 1$ (dashed line), using the same normalization. In the asymptotic limit, when $aL\gg 1$, it is found that $I_\mathrm{sp}=-I_\mathrm{st}\propto aL$ and the two intensities cancel each other exactly. 
    } 
    \label{fig:Isp}
\end{figure}

Figure \ref{fig:Isp} shows the steady-state intensity due to spontaneous emission $I_\mathrm{sp}$ (blue curve) as a function of $aL$, as calculated with equations (\ref{eq:Isp(L)_main}) and (\ref{eq:C(z,z')_steady_main}) using these parameters. Also shown is $-I_\mathrm{st}$, the negative of the stimulated intensity (red line). We can verify that, as previously discussed, $I_\mathrm{sp}=-I_\mathrm{st}\propto L$ in the astrophysical asymptotic limit when $aL\gg 1$, resulting in the cancellation of the two intensities. We therefore find that such conditions are conducive to establishing subradiance at a level that renders atomic hydrogen dark in emission and completely transparent to an incident non-coherent radiation field.

A few more features apparent in the figure are worth discussing. We first revisit the asymptotic limit results discussed in Sec.~\ref{sec:steady-state} to compare the spontaneous intensity level of equation (\ref{eq:Isp_asym_main}) with that expected from a non-coherent system
\begin{align}
    I_\mathrm{nc}\left(L\right) = 2\hbar\omega_0\Gamma n'_bL,\label{eq:Inc_asym}
\end{align}
with $n'_b$ (half) the population of the upper level. This comparison thus reveals an overwhelming suppression of spontaneous intensity through subradiance, since $\beta_{k'}/\Gamma\sim 10^{-4}$ and $n'_0/n'_b\sim 10^{-3}$.

Second, contrary to the Beer-Lambert law, which leads to an exponential decay in stimulated intensity with distance in a non-coherent gas, equation (\ref{eq:Ist_steady_main}) shows that the level of attenuation sustained by the incident radiation field remains linear with distance no matter how large $L$ becomes. This strong departure from the Beer-Lambert law reveals an anomalous behaviour in the absorption properties of the gas in the asymptotic limit.  

Figure \ref{fig:Isp} reveals a second regime, i.e., the small-sample limit where $I_\mathrm{sp}\propto \left(aL\right)^2$ (dashed line), when $aL\ll 1$, also discussed in Appendix \ref{app:small-sample}. It is then shown that
\begin{align}
    I_\mathrm{sp}\left(L\right) = 2\hbar\omega_0\beta_{k'}n'_0L\left(aL\right),\label{eq:Isp(L)_small_main}
\end{align}
which, from a comparison with equation (\ref{eq:Isp_asym_main}) for the asymptotic limit, shows that the already heavily suppressed intensity is further reduced by a factor $aL\ll 1$, once again emphasizing the strong level of subradiance. The quadratic dependency on $L$ is a clear signature of the coherent cooperative emission and implies that the entire sample is phase coherent. The atoms radiate as a single coherent entity, yielding a heavily suppressed subradiance emission scaling as $\propto L^2$.

This behaviour justifies our earlier \textit{subradiance coherence length-scale} appellation for the parameter $a^{-1}$ of equation (\ref{eq:a}), since it separates the coherent regime where the whole system acts cooperatively, when $aL\ll 1$, and the astrophysical asymptotic regime where the system spans many coherence lengths. For the latter, different regions are radiating independently from one another leading to the subradiance intensity scaling $\propto L$ of equation (\ref{eq:Isp_asym_main}).  

Another important property of our darkened atomic hydrogen gas is discussed in Appendix \ref{app:cross_section}. There, we calculate the cross-section for high-speed collisions (i.e., when the kinetic energy is far greater than the transition excitation energy $\hbar\omega_0$) between two entangled atomic systems such as the ones considered here. Interestingly, it is found that in such circumstances the collision cross-section will effectively vanish because of destructive interference due to the entangled nature of the states the systems find themselves in the steady-state, i.e, when the symmetric and ground states are depleted and the bulk of the population resides in the dark edge (see Figure \ref{fig:pop}). This result is consistent with astronomical observations of the so-called bullet cluster (with collision speeds on the order of $10^3\,\mathrm{km\,s^{-1}}$) and similar objects \cite{Clowe2006}. These observations put a constraint on the ratio of the collision cross-section to the mass of a purported dark matter (point) particle to $\sigma/M\lesssim 1\,\mathrm{cm^2\,g^{-1}}$ \cite{Cirelli2024}. Although the corresponding figure for atomic hydrogen is $\sim10^9\,\mathrm{cm^2\,g^{-1}}$, it does not apply in our case. Our analysis does not suggest that atomic hydrogen is a dark matter particle, but rather that the cooperative quantum optical behaviour of entangled hydrogen gases could account for the non-luminous matter detected in galactic halos, without invoking any new physics beyond well-established quantum electrodynamics applied to the most abundant element in the universe. In other words, it is the collective quantum optical state of the gas, not the individual atoms, that we propose as a candidate for the dark matter in galactic halos. While individual atoms may collide, entangled systems of atoms will be essentially collision-less.  

The properties of HVCs in galactic halos provide additional context for our results. In particular, the thermal bandwidth of the CNM \HI gas at 100~K corresponds to $\Delta\nu\approx 4\times 10^3$~Hz, which puts a constraint on the coherence time-scale $T_R$ for the transient response of the gas to ensure that all atoms act cooperatively. That is, we require
\begin{align}
    T_R^{-1} &= \frac{3\Gamma n'_0\lambda^2L}{4\pi}\nonumber \\
    &\gtrsim \Delta\nu, \label{eq:1/TR}
\end{align}
which is satisfied for $L\sim 50$~pc for a gas density of $0.5\;\mathrm{cm}^{-3}$ (and $n'_0=8.5\times 10^{-5}\;\mathrm{cm}^{-3}$). This density, which leads to our adopted minimal atomic separation $\Delta r_\mathrm{min}=0.06\lambda$, fulfills the necessary condition for subradiance that a large number of atoms is found within a volume $\lambda^3$. However, observational findings for the velocity line widths of discrete \HI clouds around M31 suggest a dark-to-\HI mass ratio of $\sim$100:1 for gravitationally bound entities \cite{Thilker2004}. Our subradiance/dark matter mechanism implies that the true total atomic hydrogen density is $\sim 100$ times higher than the detected \HI density, bringing the minimum length-scale to $L\sim0.5$~pc, which is readily satisfied with the approximate typical size of a CNM core \cite{Marchal2021}. An increase in kinetic temperature coupled to a decrease in density, as found in the WNM, could reduce the effective density of cooperating atoms below the density threshold needed for subradiance to operate and the sound applicability of our continuum model.  The warm envelope would therefore radiate ``normally'' and accounts for the detected 21-cm emission. The cold CNM cores, however, satisfy the conditions for subradiance and the bulk of their hydrogen is rendered dark in emission, transparent to incident radiation and collision-less. 

As the 21~cm line is one of the primary observational tools used to trace neutral hydrogen in galaxies and their halos, large blind surveys have provided censuses of \HI masses for tens of thousands of galaxies in the local universe \cite{Meyer2004,Giovanelli2005,Haynes2018}. In the scenario presented in this work, subradiant hydrogen gas in galactic halos would be effectively invisible to 21~cm emission surveys, since subradiance renders the gas dark in emission and transparent to incident radiation. As a result, \HI masses derived from 21~cm emission would systematically underestimate the true neutral hydrogen content of galactic halos; precisely the discrepancy currently attributed to dark matter. 

Finally, our model makes several testable predictions. For example, the halo mass should increase with the 21~cm flux density from the host galaxy, and for cold cores in HVCs a systematic discrepancy should exist between \HI column densities inferred from 21~cm emission and those from other tracers such as Lyman-$\alpha$ absorption (since subradiance could not operate for these shorter wavelengths). We also emphasize that the mechanism proposed here operates specifically in the context of galactic halos at the present epoch, and we make no claim regarding the nature of  dark matter on cosmological scales or in the early universe, where other constraints apply and where our mechanism may not operate.


\bibliography{DM-bib}

\backmatter

\bmhead{Acknowledgements}

We thank J.~Cami, B.~Lankhaar and A.~Mathews for their helpful comments. M.H.’s research is funded through the Natural Sciences and Engineering Research Council of Canada (NSERC) Discovery Grant RGPIN-2024-05242 and the Western Strategic Support for Research Accelerator Success. F.R.'s research is supported by the NSERC Discovery Grant RGPIN-2024-06346. M.H. and F.R. are grateful for the hospitality of Perimeter Institute where part of this work was carried out. 

\section*{Author contributions:}

M.H. and F.R. conceptualized the work and contributed to the analysis and writing of the paper. L.S. and V.A. helped with computations.

\section*{Declarations}

The authors declare no competing interest.

\clearpage


\begin{center}
\vspace{0.3cm}
\textbf{Supplementary information to:}\\
\vspace{0.1cm}
\textbf{Quantum coherence and the invisible Universe: Subradiance as a dark matter mechanism}\\
\vspace{0.3cm}
Martin Houde$^{1^\ast}$, Fereshteh Rajabi$^{2}$, Lamies Sati$^{1}$, Vahid Anari$^{1,2}$\\
\vspace{0.1cm}
$^{1^\ast}$Department of Physics and Astronomy, The University of Western Ontario,\\
1151 Richmond Street, London, N6A 3K7, Ontario, Canada\\
\vspace{0.1cm}
$^{2}$Department of Physics and Astronomy, McMaster University,\\ 
1280 Main Street West, Hamilton, L8S 4L8, Ontario, Canada\\
\vspace{0.5cm}
$^\ast$Corresponding author. E-mail: \href{mailto:mhoude2@uwo.ca}{mhoude2@uwo.ca}\\

\vspace{1cm}
\textbf{CONTENTS}
\end{center}
\vspace{0.5cm}

\begin{tabular}{l r}
Appendix \ref{app:Hamiltonian}: Hamiltonian and transition rates & \pageref{app:Hamiltonian}\\
\indent Two-atom transition rates from $m=0$ and superabsorption & \pageref{app:two-atom}\\
\indent Energy trapping efficiency for $n$ atoms & \pageref{sec:n-atom}\\
\\
Appendix \ref{app:master equation}: Generalized master equation & \pageref{app:master equation}\\
\\
Appendix \ref{app:MBE}: The continuum equations & \pageref{app:MBE}\\
\indent Steady-state solution in the linear regime & \pageref{app:linear}\\
\indent \indent Asymptotic astrophysical limit ($aL\gg 1$) & \pageref{app:asymptotic}\\
\indent \indent Small-sample limit ($aL\ll 1$) & \pageref{app:small-sample}\\
\\
Appendix \ref{app:cross_section}: Collision cross-section & \pageref{app:cross_section}\\
\indent Generalization to arbitrarily large atomic ensembles & \pageref{sec:N_ensembles} \\
\indent Independence on internal states for high energy scattering & \pageref{app:independence}
\end{tabular}

\clearpage

\begin{appendices}

\section{Hamiltonian and transition rates}\label{app:Hamiltonian}

Following Dicke \cite{Dicke1954}, we express the Hamiltonian for the gas with 
\begin{equation}
    \hat{H} = \hat{H}_0 + \hbar\omega_0\sum_j\hat{R}^3_j + \hat{V}, \label{eq:H}
\end{equation}
\noindent where $\hat{H}_0$ accounts for the translational motions of the atoms' center-of-mass, the second term for their internal energy and $\hat{V}$ for dipole interactions with the electromagnetic field. For simplicity we approximate the atoms as two-level systems at resonance with an internal energy difference $\hbar\omega_0$; the summation in equation (\ref{eq:H}) is on all atoms. That is, if $\ket{a}$ and $\ket{b}$ are the lower and upper atomic states, respectively, then for the $j^\mathrm{th}$ atom $\hat{R}^3_j = \frac{1}{2}\left(\Ketbra{b}{b}-\Ketbra{a}{a}\right)$.

For our problem, the interaction term reduces to
\begin{equation}
    \hat{V} = -i\sum_{k} \frac{\hbar}{2}\Omega_{k}\left(\hat{R}^+_{k}\hat{a}_{k} - \hat{R}^-_{k}\hat{a}^\dagger_{k}\right), \label{eq:V} 
\end{equation}
\noindent where the summation on $k$ is on the radiation modes, and $\hat{a}_{k}$ and $\hat{a}^\dagger_{k}$ are the photon annihilation and creation operators. In equation (\ref{eq:V}) the Rabi frequency is given by
\begin{equation}
    \Omega_{k} = \frac{2\mu\mathcal{E}_0}{\hbar c}\left(\boldsymbol{\epsilon}_\mu\cdot\boldsymbol{\epsilon}_{k}^{\perp}\right) \label{eq:Omega_k}
\end{equation}
\noindent for a magnetic dipole transition, where $\mu$ and $\boldsymbol{\epsilon}_\mu$ are the magnitude of the magnetic dipole moment and its associated unit vector (assumed real), while that for the radiation mode's polarization is (also real and) denoted by $\boldsymbol{\epsilon}_{k}^\perp$ (i.e., $\mathbf{k}=k\boldsymbol{\epsilon}_{k}$ and $\boldsymbol{\epsilon}_{k}\cdot\boldsymbol{\epsilon}_{k}^\perp=0$). The one-photon electric field is
\begin{equation}
\mathcal{E}_0 = \sqrt{\frac{\hbar\omega_0}{2\epsilon_0 \mathcal{V}}} \label{eq:one-photon}
\end{equation}
\noindent with $\epsilon_0$ the permittivity of vacuum and $\mathcal{V}$ the volume of quantization. We focus on magnetic dipole transitions in view of our application to the atomic hydrogen 21~cm line in Sec.~\ref{sec:global entanglement}. For an electrical dipole transition one has to substitute $\mu\rightarrow dc$ and $\boldsymbol{\epsilon}_\mu\rightarrow \boldsymbol{\epsilon}_d$ in equation (\ref{eq:Omega_k}), with $d$ the magnitude of the electric dipole moment.

The phase-matched raising and lowering operators introduced in equation (\ref{eq:V}) are defined by
\begin{equation}
    \hat{R}^\pm_{k} = \sum_j\hat{R}^\pm_j e^{\pm i\mathbf{k}\cdot\mathbf{r}_j} \label{eq:R+-}
\end{equation}
\noindent  with $\mathbf{r}_j$ the position of atom $j$. For the single-atom raising and lowering operators we have $\hat{R}^+_j = \Ketbra{b}{a}$ and $\hat{R}^-_j = \Ketbra{a}{b}$ while, as we assume resonance, we write $k=\left|\mathbf{k}\right|=\omega_0/c$. For two identical two-level atoms located a distance $z_0$ from one another (Atom 1 at $-z_0/2$ and Atom 2 at $z_0/2$ on the $z$-axis), as considered in Sec.~\ref{sec:entangled state}, we have 
\begin{equation}
    \hat{R}^\pm_{k} = \hat{R}^\pm_1 e^{\mp i\frac{1}{2}k z_0 \cos\theta} + \hat{R}^\pm_2 e^{\pm i\frac{1}{2}k z_0 \cos\theta} \label{eq:R+-_two}
\end{equation}
\noindent with $\theta$ the angle of $\textbf{k}$ relative to the $z$-axis.

\subsection{Two-atom transition rates from $m=0$}\label{app:two-atom}

Starting with equations (\ref{eq:10->1-1})-(\ref{eq:00->1-1}), we can obtain the radiation intensity by integrating these transition rates over $d\Omega^\prime$. Setting $\cos\theta=0$, for simplicity, we find for the transition rates
\begin{align}
    & \gamma_{1,0_\theta\rightarrow 1,-1} =  \Gamma\left[1+F\left(kz_0\right)\right] \label{eq:g10->1-1_app} \\
    & \gamma_{0,0_\theta\rightarrow 1,-1} = \Gamma\left[1-F\left(kz_0\right)\right] \label{eq:g00->1-1_app}
\end{align}
with
\begin{align}
    F\left(kz_0\right) = & \frac{3}{2}\left\{\rule{0mm}{6mm}\left[1-\left(\boldsymbol{\epsilon}_{\mu}\cdot\boldsymbol{\epsilon}_{z}\right)^2\right]\frac{\sin\left(kz_0\right)}{kz_0}\right.\nonumber\\
    & \left. +\left[1-3\left(\boldsymbol{\epsilon}_{\mu}\cdot\boldsymbol{\epsilon}_{z}\right)^2\right]\left[\frac{\cos\left(kz_0\right)}{\left(kz_0\right)^2}-\frac{\sin\left(kz_0\right)}{\left(kz_0\right)^3}\right]\right\} \label{eq:F}
\end{align}
and
\begin{align}
    \Gamma=\frac{\mu_0\omega_0^3 \mu^2}{3\pi\hbar c^3} \label{eq:Gamma}
\end{align}
the single-atom free-space spontaneous emission rate with $\mu_0$ the permeability of vacuum (for an electric dipole transition we again substitute $\mu\rightarrow dc$). 

We consider initial conditions where the states $\Ket{1,0}_\theta$ and $\Ket{0,0}_\theta$ have equal probabilities of occupation (of $1/2$) but with uncorrelated probability amplitude coefficients. Under these conditions, the radiation intensity
can be evaluated with 
\begin{align}
    I\left(t\right) = \hbar\omega_0\left[P_{1,0_\theta}\left(t\right)\gamma_{1,0_\theta\rightarrow 1,-1}+P_{0,0_\theta}\left(t\right)\gamma_{0,0_\theta\rightarrow 1,-1}\right], \label{eq:I}
\end{align}
where $P_{r,m_\theta}\left(t\right)$ is the probability of being in state $\Ket{r,m}_\theta$ at time $t$; the different $P_{r,m_\theta}\left(t\right)$ are readily calculated from equations (\ref{eq:g10->1-1_app})-(\ref{eq:g00->1-1_app}) and the initial probabilities. We thus have 
\begin{align}
    I\left(t\right) = \frac{1}{2}\hbar\omega_0\Gamma e^{-\Gamma t}&\left\{ \left[1+F\left(kz_0\right)\right]e^{-\Gamma F\left(kz_0\right)t} + \left[1-F\left(kz_0\right)\right]e^{\Gamma F\left(kz_0\right)t}\right\}. \label{eq:T_solved}
\end{align}
The two limits given in equations (\ref{eq:I_inf})-(\ref{eq:I_ss}) result from this relation with $F\left(0\right)=1$ and $F\left(\infty\right)=0$.

The transition rates for super- and subabsorption are obtained when the two-atom system is initially in the ground $\Ket{1,-1}$ state and subjected to an incident one-photon field. It is then found that the transition rates of equations (\ref{eq:11->10})-(\ref{eq:00->1-1}) are reversed \cite{Yang2021}. That is,
\begin{align}
    \frac{d\gamma_{1,-1\rightarrow 1,0_\theta}}{d\Omega} &= 2\frac{d\Gamma}{d\Omega} \label{eq:1-1->10} \\
    \frac{d\gamma_{1,-1\rightarrow 0,0_\theta}}{d\Omega} &= 0 \label{eq:1-1->00} \\
    \frac{d\gamma_{1,0_\theta\rightarrow 1,1}}{d\Omega^\prime} &= 2\cos^2\left[\frac{1}{2}k z_0\left(\cos\theta^\prime-\cos\theta\right)\right]\frac{d\Gamma}{d\Omega^\prime} \label{eq:10->11} \\
    \frac{d\gamma_{0,0_\theta\rightarrow 1,1}}{d\Omega^\prime} &= 2\sin^2\left[\frac{1}{2}k z_0\left(\cos\theta^\prime-\cos\theta\right)\right]\frac{d\Gamma}{d\Omega^\prime} \label{eq:00->11} 
\end{align}
with $\theta$ and $\theta^\prime$ denoting the orientations of the first and second absorbed photons, respectively, relative to the $z$-axis. Here again, there is angular correlation, but this time between successively absorbed photons.

\subsection{Energy trapping efficiency for $n$ atoms}\label{sec:n-atom}

For the two-atom problem only two permutation symmetries are available for the Dicke states: the totally symmetric species is shared by the $r=1$ triplet (equations \ref{eq:|1,1>}-\ref{eq:|1,-1>}), while the lone $r=0$ singlet is anti-symmetric (equation \ref{eq:|0,0>}). In the more general case where the gas is composed of an arbitrary number of atoms $n$ a larger number of symmetries, tied to the cooperative number $r$, exist for the Dicke states (see Fig.~\ref{fig:Dicke-table}). 

Although it is in principle possible to extend the spontaneous emission rates analysis for finite separations performed for the two-atom case to the $n$-atom problem using appropriate Dicke states, the calculations quickly become prohibitive even for a restricted number of radiators. It is, however, possible to determine the ratio of numbers of photons emitted in the small and infinite-size samples when $m=0$ initially, as done in equation (\ref{eq:eta}) for the two-atom problem. For this, we use the known level of degeneracy of states as a function of the cooperative number $r$ \cite{Dicke1954}
\begin{equation}
    g_r = \frac{n!\left(2r+1\right)}{\left(n/2+r+1\right)!\left(n/2-r\right)!},\label{eq:g_r}
\end{equation}
\noindent with the fact that, for $n$ even, the number of photons emitted from $m=0$ for the small and infinite-size samples are $r$ and $n/2$, respectively. We thus find for the energy trapping efficiency 
\begin{equation}
    \eta = 1-\frac{2}{n}\frac{\sum_{r=0}^{n/2} g_r r}{\sum_{r=0}^{n/2} g_r}.\label{eq:eta_n}
\end{equation}
\noindent One can easily verify that $\eta=1/2$ for the two-atom case, while it is $0.69$ for $n=10$, $0.88$ for $n=100$, etc. Furthermore, it is known that $\overline{r\left(r+1\right)} \simeq m^2+n/2$ whenever $n\gg 1$ and $\hbar\omega_0\ll k_\mathrm{B}T_\mathrm{kin}$ \cite{Dicke1954}, which by equation (\ref{eq:eta_n}) yields $\eta\approx 1-\sqrt{2/n}$ when $m=0$.

\section{Generalized master equation}\label{app:master equation}

The master equation used for the analysis presented in Sect.~\ref{sec:master equation} is a generalization of what is often used for the study of superradiance \cite{Lehmberg1970,Gross1982}. That is, instead of modeling the environment with the radiation vacuum state, we allow for the existence of one non-empty radiation state $\ket{n_{k^\prime}}$ containing $n_{k^\prime}$ photons of mode $\mathbf{k}^\prime=\left(\omega_0/c\right)\boldsymbol{\epsilon}_{k^\prime}$ (i.e., $\left|\mathbf{k}^\prime\right|=k$) with linear polarization $\boldsymbol{\epsilon}_{k^\prime}^\perp$ ($\boldsymbol{\epsilon}_{k^\prime}\cdot\boldsymbol{\epsilon}_{k^\prime}^\perp=0$), while all the other modes are assumed empty. This allows us to account for the existence of a non-coherent incident radiation field on the atomic ensemble under study. This generalized master equation takes the form
\begin{align}
    \frac{d\hat{\rho}}{dt} &= \frac{\omega_0}{i}\sum_j\left[\hat{R}^3_j,\hat{\rho}\right]-\frac{\Gamma}{i}\sum_{i\neq j}\Omega_{ij}\left(k\Delta r_{ij}\right)\left[\hat{R}^+_i\hat{R}^-_j,\hat{\rho}\right]\nonumber \\
    & -\frac{\Gamma}{2}\sum_{ij}F_{ij}\left(k\Delta r_{ij}\right)\left(\hat{R}^+_i\hat{R}^-_j\hat{\rho}+\hat{\rho}\hat{R}^+_i\hat{R}^-_j-2\hat{R}^-_j\hat{\rho}\hat{R}^+_i\right)\nonumber \\
    & -\frac{\beta_{k^\prime}}{2}\sum_{ij}\left[\left(\hat{R}^+_{k^\prime i}\hat{R}^-_{k^\prime j}\hat{\rho}+\hat{\rho}\hat{R}^+_{k^\prime i}\hat{R}^-_{k^\prime j}-2\hat{R}^-_{k^\prime j}\hat{\rho}\hat{R}^+_{k^\prime i}\right)\right.\nonumber \\
    & \qquad\qquad\left.+\left(\hat{R}^-_{k^\prime i}\hat{R}^+_{k^\prime j}\hat{\rho}+\hat{\rho}\hat{R}^-_{k^\prime i}\hat{R}^+_{k^\prime j}-2\hat{R}^+_{k^\prime j}\hat{\rho}\hat{R}^-_{k^\prime i}\right)\right] \label{eq:master}
\end{align}
with
\begin{align}
    F_{ij}\left(k\Delta r_{ij}\right) & = \frac{3}{2}\left\{\rule{0mm}{6mm}\left[1-\left(\boldsymbol{\epsilon}_{\mu}\cdot\boldsymbol{\epsilon}_{r}\right)^2\right]\frac{\sin\left(k\Delta r_{ij}\right)}{k\Delta r_{ij}}\right.\nonumber\\
    & \left. +\left[1-3\left(\boldsymbol{\epsilon}_{\mu}\cdot\boldsymbol{\epsilon}_{r}\right)^2\right]\left[\frac{\cos\left(k\Delta r_{ij}\right)}{\left(k\Delta r_{ij}\right)^2}-\frac{\sin\left(k\Delta r_{ij}\right)}{\left(k\Delta r_{ij}\right)^3}\right]\right\} \label{eq:Fij}\\
    \Omega_{ij}\left(k\Delta r_{ij}\right) & = \frac{3}{4}\left\{\rule{0mm}{6mm}\left[1-\left(\boldsymbol{\epsilon}_{\mu}\cdot\boldsymbol{\epsilon}_{r}\right)^2\right]\frac{\cos\left(k\Delta r_{ij}\right)}{k\Delta r_{ij}}\right.\nonumber\\
    & \left. -\left[1-3\left(\boldsymbol{\epsilon}_{\mu}\cdot\boldsymbol{\epsilon}_{r}\right)^2\right]\left[\frac{\sin\left(k\Delta r_{ij}\right)}{\left(k\Delta r_{ij}\right)^2}+\frac{\cos\left(k\Delta r_{ij}\right)}{\left(k\Delta r_{ij}\right)^3}\right]\right\} \label{eq:Omegaij}\\
    \beta_{k^\prime} & = n_{k^\prime}\left(\frac{3\Gamma}{8\pi}\cdot\frac{\lambda}{L}\right)\left(\boldsymbol{\epsilon}_{\mu}\cdot\boldsymbol{\epsilon}_{k^\prime}^\perp\right)^2 \label{eq:beta_k} \\
    \hat{R}^\pm_{k^\prime j} & = \hat{R}^\pm_je^{\pm i\mathbf{k}^\prime\cdot\mathbf{r}_j} \label{eq:R+-_kj}
\end{align}
and where $\hat\rho$ is the density matrix for the atomic system obtained by taking the trace of its radiation counterpart over all possible radiation modes, $\Delta r_{ij}=\left|\mathbf{r}_i-\mathbf{r}_j\right|$ is the distance between atoms $i$ and $j$, $\boldsymbol{\epsilon}_{r}=\left(\mathbf{r}_i-\mathbf{r}_j\right)/\Delta r_{ij}$ and $\Gamma$ is given by equation (\ref{eq:Gamma}). In all calculations presented in Sec.~\ref{sec:master equation} we set $\left(\boldsymbol{\epsilon}_{\mu}\cdot\boldsymbol{\epsilon}_{r}\right)^2=1/3$, which corresponds to a random alignment between the two vectors, and $\left(\boldsymbol{\epsilon}_{\mu}\cdot\boldsymbol{\epsilon}_{k^\prime}^\perp\right)^2=1$ for simplicity. The first two lines in equation (\ref{eq:master}) make up the usual master equation for the study of (cooperative) spontaneous emission in the vacuum \citep{Lehmberg1970}, while the third and fourth lines are, respectively, for (cooperative) stimulated emission and absorption stemming from the presence of the incident radiation field of mode $\ket{n_{k^\prime}}$.

Importantly for our analysis, the intensities (in units of energy per second) due to spontaneous emission and stimulated processes (emission and absorption) are respectively given by
\begin{align}
    I_\mathrm{sp} &= \hbar\omega_0\Gamma\sum_{ij}F_{ij}\left(k\Delta r_{ij}\right)\left<\hat{R}^+_i\hat{R}^-_j\right>\label{eq:Isp-coop}\\
    &= \hbar\omega_0\Gamma\left[\sum_i\left(\frac{1}{2}+\left<\hat{R}^3_i\right>\right)+\sum_{i\neq j}F_{ij}\left(k\Delta r_{ij}\right)\left<\hat{R}^+_i\hat{R}^-_j\right>\right]\label{eq:Isp-R3}\\
    I_\mathrm{st} &= \hbar\omega_0\beta_{k^\prime}\sum_{ij}\left<\hat{R}^+_{k^\prime i}\hat{R}^-_{k^\prime j}-\hat{R}^-_{k^\prime i}\hat{R}^+_{k^\prime j}\right>\nonumber\\
    &= 2\hbar\omega_0\beta_{k^\prime}\sum_i\left<\hat{R}^3_i\right>,\label{eq:Ist-R3}
\end{align}
where $\left<\hat{X}\right>\equiv\mathrm{Tr}\left\{\hat{X}\hat{\rho}\right\}$. Equation (\ref{eq:Isp-R3}) for the spontaneous intensity is particularly informative as it separates the non-coherent single-atom (first term) and cooperative (involving pairs of atoms; second term) contributions to the overall intensity. The cooperative term can enhance or suppress the level of intensity, while in the absence of interatomic correlations (e.g., when $\Delta r_{ij}\gg\lambda$) the spontaneous emission tends to the non-coherent contributions of the independent atoms populating the upper state.

When numerically solving equation (\ref{eq:master}), we add phenomenological relaxation and dephasing terms operating on time-scales $T_1$ and $T_2$, respectively, to account for processes that can hamper coherence in the system (e.g., collisions). A pump, also operating on the $T_1$ time-scale, is further inserted to set equilibrium conditions.  

We can illustrate this by considering the time evolution of the two-atom system, with the atoms located at $\pm \left(z_0/2\right)\,\mathbf{e}_z$ and $\mathbf{k}^\prime\cdot\mathbf{e}_z=k\cos\theta$. Using the corresponding atomic states defined in equations (\ref{eq:|1,1>})-(\ref{eq:|0,0>}), the temporal evolution of the density matrix elements $\rho_{00}\equiv \braket{1,-1|\hat{\rho}|1,-1}$, $\rho_{11}\equiv \left._\theta\braket{1,0|\hat{\rho}|1,0}_\theta\right.$, $\rho_{22}\equiv \left._\theta\braket{0,0|\hat{\rho}|0,0}_\theta\right.$, $\rho_{33}\equiv \braket{1,1|\hat{\rho}|1,1}$ and $\rho_{12}\equiv \left._\theta\braket{1,0|\hat{\rho}|0,0}_\theta\right.$ can be calculated from equations (\ref{eq:master})-(\ref{eq:R+-_kj}), yielding the following set of closed relations 
\begin{align}
    \frac{1}{\Gamma}\frac{d\rho_{00}}{dt} &= -\frac{2\beta_{k^\prime}}{\Gamma}\rho_{00}+\left(1+\gamma\cos\phi+\frac{2\beta_{k^\prime}}{\Gamma}\right)\rho_{11}+\left(1-\gamma\cos\phi\right)\rho_{22}\nonumber\\
    &-i\gamma\sin\phi\left(\rho_{12}-\rho_{21}\right) -\frac{\left(\rho_{00}-\rho_{00}^\mathrm{eq}\right)}{\Gamma T_1} \label{eq:dp00/dt}\\
    \frac{1}{\Gamma}\frac{d\rho_{11}}{dt} &= -\left(1+\gamma\cos\phi+\frac{4\beta_{k^\prime}}{\Gamma}\right)\rho_{11}+\left(1+\gamma\cos\phi+\frac{2\beta_{k^\prime}}{\Gamma}\right)\rho_{33}+\frac{2\beta_{k^\prime}}{\Gamma}\rho_{00}\nonumber\\
    & -\Omega\sin\phi\left(\rho_{12}+\rho_{21}\right)+i\frac{\gamma}{2}\sin\phi\left(\rho_{12}-\rho_{21}\right)-\frac{\left(\rho_{11}-\rho_{11}^\mathrm{eq}\right)}{\Gamma T_1}\label{eq:dp11/dt}\\
    \frac{1}{\Gamma}\frac{d\rho_{22}}{dt} &= -\left(1-\gamma\cos\phi\right)\left(\rho_{22}-\rho_{33}\right)+\Omega\sin\phi\left(\rho_{12}+\rho_{21}\right)+i\frac{\gamma}{2}\sin\phi\left(\rho_{12}-\rho_{21}\right)\nonumber\\
    & -\frac{\left(\rho_{22}-\rho_{22}^\mathrm{eq}\right)}{\Gamma T_1}\label{eq:dp22/dt}\\
    \frac{1}{\Gamma}\frac{d\rho_{33}}{dt} &= -2\left(1+\frac{\beta_{k^\prime}}{\Gamma}\right)\rho_{33}+\frac{2\beta_{k^\prime}}{\Gamma}\rho_{11}-\frac{\left(\rho_{33}-\rho_{33}^\mathrm{eq}\right)}{\Gamma T_1}\label{eq:dp33/dt}\\
    \frac{1}{\Gamma}\frac{d\rho_{12}}{dt} &= -\left(1+\frac{2\beta_{k^\prime}}{\Gamma}\right)\rho_{12}+i2\Omega\cos\phi\rho_{12}\nonumber\\
    & +\Omega\sin\phi\left(\rho_{11}-\rho_{22}\right)+i\frac{\gamma}{2}\sin\phi\left(2\rho_{33}-\rho_{11}-\rho_{22}\right)-\frac{\rho_{12}}{\Gamma T_2},\label{eq:dp12/dt}
\end{align}
where $\gamma=F_{12}\left(k\Delta z_0\right)$, $\Omega=\Omega_{12}\left(k\Delta z_0\right)$ and $\phi=k\Delta z_0\cos\theta$. The relaxation and pumping phenomenological terms are explicit in the evolution equations for the populations (i.e., equations \ref{eq:dp00/dt}-\ref{eq:dp33/dt}) while dephasing is applied to coherences (as in equation \ref{eq:dp12/dt}).

A closer look at equations (\ref{eq:dp00/dt})-(\ref{eq:dp33/dt}) reveals that the pump terms are responsible for finite, non-zero, population levels in the steady-state. More precisely, setting the time-derivatives to zero and $\phi=\beta_{k'}=\rho_{00}^\mathrm{eq}=\rho_{33}^\mathrm{eq}=0$, for simplicity, yields the steady-state populations
\begin{align}
    \rho_{11}^\mathrm{ss} &= \frac{\rho_{11}^\mathrm{eq}}{1+\Gamma T_1\left(1+\gamma\right)}\label{eq:p11^ss}\\
    \rho_{22}^\mathrm{ss} &= \frac{\rho_{22}^\mathrm{eq}}{1+\Gamma T_1\left(1-\gamma\right)},\label{eq:p22^ss}
\end{align}
for example. In general, we find a lower steady-state population level for a higher coherent decay rate, and vice-versa. 

Finally, we note that while $\beta_{k'}$ can readily be related to the Einstein stimulated emission coefficient, it can also be shown that, when considering the radiation flux density $F_\nu$ irradiated in a halo from a galactic disk, 
\begin{align}
    \frac{\beta_{k'}}{\Gamma} = \frac{3\lambda^3}{8\pi^2\hbar\omega_0c}F_\nu\Delta\nu\label{eq:beta_halo}
\end{align}
with $\Delta\nu$ the spectral bandwidth of the radiation field ($\omega_0=2\pi\nu$). 

An estimate for the continuum radiation field at $\lambda=21$~cm can be obtained from the corresponding total luminosity of a typical star-forming galaxy. Using a luminosity $L_\nu\left(1.4\;\mathrm{GHz}\right)\sim 10^{22}-10^{23}\;\mathrm{W\;Hz^{-1}}$ \cite{Mobasher1999}, the flux density at a distance $D\sim 10$~kpc above or below the galactic disk yields
\begin{align}
    F_\nu\left(1.4\;\mathrm{GHz}\right) = \frac{L_\nu}{4\pi D^2}\sim 10^6\;\mathrm{Jy}.\label{eq:Fnu} 
\end{align}
We therefore find $\beta_{k'}/\Gamma\sim 10^{-4}$ for $\Delta\nu\sim 10^4$~Hz (i.e., for a kinetic temperature of a few times $100$~K).  

\section{The continuum equations}\label{app:MBE}

For systems containing a large number of atoms, the discrete master equation (\ref{eq:master}) can be transformed to a continuum limit by defining coarse-grained densities for the pseudo-spin operators at a given position $\mathbf{r}$ 
\begin{align}
    \hat{R}^\alpha\left(\mathbf{r},t\right) = \frac{1}{\delta \mathcal{V}}\sum_{j\in\delta \mathcal{V}} \hat{R}_j^\alpha,\quad\mathrm{for}\;\alpha=+,-\;\mathrm{and}\;3,\label{eq:Ralpha}
\end{align}
and replacing summations by integrals. In equation (\ref{eq:Ralpha}), it is assumed that the volume $\delta \mathcal{V}$, centred at $\mathbf{r}$, contains a sufficient number of atoms to ensure proper statistics while being much smaller than the wavelength \citep{Gross1982}. Using the generalized master equation with $d\left<\hat{X}\right>/dt=\mathrm{Tr}\left\{\hat{X}\,d\hat{\rho}/dt\right\}$ and approximating $\left<\hat{A}\left(\mathbf{r}\right)\hat{R}^3\left(\mathbf{r}'\right)\hat{B}\left(\mathbf{r}''\right)\right>=\left<\hat{R}^3\left(\mathbf{r}'\right)\right>\left<\hat{A}\left(\mathbf{r}\right)\hat{B}\left(\mathbf{r}''\right)\right>$, etc. \cite{Park2024}, we obtain the following closed set of equations 
\begin{align}
    \frac{dn^\prime\left(\mathbf{r},t\right)}{dt} &= -\Gamma\int d^3r'\,\mathrm{Re}\left[G_\mathrm{fs}\left(k\left|\mathbf{r}-\mathbf{r}^\prime\right|\right)C^\ast\left(\mathbf{r},\mathbf{r}^\prime;t\right)\right]-2\beta_{k^\prime}n^\prime\left(\mathbf{r},t\right)\label{eq:dn'(r)}\\
    \frac{dC\left(\mathbf{r},\mathbf{r}';t\right)}{dt} &= \Gamma \left[n^\prime\left(\mathbf{r}',t\right)\int d^3r''\, G_\mathrm{fs}^*\left(k\left|\mathbf{r}'-\mathbf{r}''\right|\right)C\left(\mathbf{r},\mathbf{r}'';t\right)\right.\nonumber \\
    &+ \left. n^\prime\left(\mathbf{r},t\right)\int d^3r''\, G_\mathrm{fs}\left(k\left|\mathbf{r}''-\mathbf{r}\right|\right)C\left(\mathbf{r}'',\mathbf{r}';t\right)\right]\nonumber \\
    & - 2\beta_{k^\prime}C\left(\mathbf{r},\mathbf{r}';t\right)+4\beta_{k^\prime}e^{i\mathrm{k}'\cdot\left(\mathbf{r}-\mathbf{r}'\right)}n^\prime\left(\mathbf{r},t\right)n^\prime\left(\mathbf{r}',t\right)\label{eq:dC'(r)}
\end{align}
with $n^\prime\left(\mathbf{r},t\right)=\left<\hat{R}^3\left(\mathbf{r},t\right)\right>$ (half of) the population inversion, the two-point correlation $C\left(\mathbf{r},\mathbf{r}';t\right)=\left<\hat{R}^+\left(\mathbf{r},t\right)\hat{R}^-\left(\mathbf{r}',t\right)\right>$. The free-space Green's function is given by
\begin{align}
    G_\mathrm{fs}\left(x\right) = \frac{3}{2}\frac{e^{-ix}}{x}\left[\left(i+\frac{1}{x}-\frac{i}{x^2}\right)-\left(\boldsymbol{\epsilon}_{\mu}\cdot\boldsymbol{\epsilon}_{r}\right)^2\left(i+\frac{3}{x}-\frac{3i}{x^2}\right)\right]\label{eq:Gfs}
\end{align}
and explicitly accounts for non-local atomic correlations.

Considering a slab of thickness $L$ along the $z$-axis, we find that the Green's function simplifies as follows
\begin{align}
    G_\mathrm{slab}\left(k\left|z''-z\right|\right) &= \int_{-\infty}^\infty dx''dy''\, G_\mathrm{fs}\left(k\left|\mathbf{r}''-\mathbf{r}\right|\right)\nonumber \\
    &= \frac{3\pi}{k^2}e^{-ik\left|z''-z\right|}.\label{eq:slab}
\end{align}
For the astrophysical systems considered in this paper (where $L\gg\lambda$), this slab Green's function only admits propagating eigenmodes such that $C\left(z'',z;t\right)=\tilde{C}\left(z'',z;t\right)e^{ik\left(z-z''\right)}$. It follows that equations (\ref{eq:dn'(r)})-(\ref{eq:dC'(r)}) reduce to
\begin{align}
    \frac{dn^\prime\left(z,t\right)}{dt} &= -\frac{3\pi\Gamma}{k^2}\int_0^z dz'\,\mathrm{Re}\left[\tilde{C}\left(z,z^\prime;t\right)\right]-\left(2\beta_{k^\prime}+\frac{1}{T_1}\right)n^\prime\left(z,t\right)+\frac{n'_\mathrm{eq}}{T_1}\label{eq:dn(z)}\\
    \frac{d\tilde{C}\left(z,z';t\right)}{dt} &= \frac{3\pi\Gamma}{k^2}\left[n^\prime\left(z',t\right)\int_0^{z'} dz''\,\tilde{C}\left(z,z'';t\right)+n^\prime\left(z,t\right)\int_0^{z} dz''\,\tilde{C}\left(z'',z';t\right)\right]\nonumber \\
    &-2\left(\beta_{k'}+\frac{1}{T_2}\right)\tilde{C}\left(z,z';t\right)+4\beta_{k^\prime}n^\prime\left(z,t\right)n^\prime\left(z',t\right),\label{eq:dC(z,z')}
\end{align}
where, as for the generalized master equation, we added a phenomenological dephasing term of time-scale $T_2/2$ for the $\tilde{C}\left(z,z';t\right)$ equation, as well as a corresponding relaxation term of time-scale $T_1$ and a pump $n'_\mathrm{eq}/T_1$ for the $n'\left(z,t\right)$ equation. The function $\tilde{C}\left(z,z';t\right)$ is therefore a slowly varying envelope, which evolves on much longer scales than $\lambda$ and $\omega_0^{-1}$.

It is important to emphasize that, while the direction of propagation of the non-coherent incident field (i.e., along the $z$-axis) sets the symmetry axis of the slab, spontaneous emission into all other radiation modes is explicitly accounted for in the generalized master equation that serves as the basis for our continuum equations (see Appendix \ref{app:master equation}). Specifically, the single-atom decay rate $\Gamma$ and the cooperative kernels $F_{ij}$ and $\Omega_{ij}$, which appear in the master equation (see equations \ref{eq:master}-\ref{eq:Omegaij}) and lead to the free-space Green's function $G_\mathrm{fs}$ (see equation \ref{eq:Gfs}), already encode emission into the full solid angle of vacuum modes. The reduction to the Green's function $G_\mathrm{slab}$ (equation \ref{eq:slab}), with a corresponding propagation along the $z$-axis, and the slowly varying envelope that follows from the slab geometry is a statement about which modes are phase-matched over the length $L\gg \lambda$ and therefore dominate the cooperative dynamics, not a truncation of the radiation field.

Finally, starting from equations (\ref{eq:Isp-coop}) and (\ref{eq:Ist-R3}), the continuum spontaneous and stimulated intensities at the slab's exit become (in units of energy per second per area)
\begin{align}
    I_\mathrm{sp}\left(L,t\right) &= \hbar\omega_0\frac{3\pi\Gamma}{2k^2}\int_0^Ldz\int_0^Ldz'\,\tilde{C}\left(z,z';t\right)\label{eq:Isp(L)}\\
    I_\mathrm{st}\left(L,t\right) &= 2\hbar\omega_0\beta_{k'}\int_0^Ldz\,n'\left(z,t\right).\label{eq:Ist(L)}
\end{align}

\subsection{Steady-state solution in the linear regime}\label{app:linear}

We now focus on situations where the system is at steady-state in the linear regime, i.e., when $dn'/dt=d\tilde{C}/dt=0$ and the first term on the right-hand side in equation (\ref{eq:dn(z)}) can be neglected. We then find that the population inversion becomes constant at $n'_\mathrm{eq}/\left(1+2\beta_{k'}T_1\right)$. Since $n'_\mathrm{eq}<0$ at thermal equilibrium, it will be advantageous to introduce the steady-state density
\begin{align}
    n'_0 = \frac{\left|n'_\mathrm{eq}\right|}{1+2\beta_{k'}T_1}>0.\label{ref:n'0} 
\end{align}
Equation (\ref{eq:dC(z,z')}) can then be solved using Laplace transforms, with the $z\leftrightarrow u$ and $z'\leftrightarrow v$ correspondences, leading to 
\begin{align}
    \tilde{C}\left(u,v\right) = \frac{2n'^2_0\beta_{k'}/\alpha_{k'}}{\left(v+a\right)\left[u+va/\left(v+a\right)\right]}\label{eq:C(u,v)}
\end{align}
with $a=3\pi n'_0\Gamma/\left(2k^2\alpha_{k'}\right)$ and $\alpha_{k'}=\beta_{k'}+1/T_2$.

Effecting a first inverse Laplace transform to recover $z$ and then a second for $z'$, and using
\begin{equation}
    \mathcal{L}^{-1}_{z'}\left\{\frac{e^{b/\left(v+a\right)}}{\left(v+a\right)^n}\right\} = e^{-az'}\left(\frac{z'}{b}\right)^\frac{n-1}{2}I_{n-1}\left(2\sqrt{bz'}\right)\label{eq:BesselJ}
\end{equation}
with $I_n\left(x\right)$ the modified Bessel function of the first kind and order $n$ \cite{Abramowitz1972},
we find
\begin{align}
    \tilde{C}\left(z,z'\right) = \frac{2n'^2_0\beta_{k'}}{\alpha_{k'}}e^{-a\left(z+z'\right)}I_0\left(2a\sqrt{zz'}\right).\label{eq:C(z,z')_steady}  
\end{align}

\subsubsection{Asymptotic astrophysical limit ($aL\gg 1$)}\label{app:asymptotic}

As discussed in Secs.~\ref{sec:steady-state} and \ref{sec:global entanglement}, the asymptotic limit $aL\gg 1$ applies to astrophysical environments such as galactic halos. Because $\lim_{x\gg 1} I_0\left(x\right) = e^x/\sqrt{2\pi x}$ \cite{Abramowitz1972} we have 
\begin{align}
    \tilde{C}\left(z,z'\right) = \frac{2n'^2_0\beta_{k'}}{\alpha_{k'}}\frac{e^{-a\left(\sqrt{z}-\sqrt{z'}\right)^2}}{\sqrt{4\pi a\sqrt{zz'}}}\label{eq:C(z,z')_asym}
\end{align}
and, from equation (\ref{eq:Isp(L)}),
\begin{align}
    I_\mathrm{sp}\left(L\right) = 2\hbar\omega_0\beta_{k'}n'_0 L.\label{eq:Isp_asym}
\end{align}

\subsubsection{Small-sample limit ($aL\ll 1$)}\label{app:small-sample}

It is also interesting to consider the opposite, small-sample, limit when $aL\ll 1$ and $\lim_{x\ll 1}I_0\left(x\right)=1$ \cite{Abramowitz1972}. We then have 
\begin{align}
    \tilde{C}\left(z,z'\right) = \frac{2n'^2_0\beta_{k'}}{\alpha_{k'}}e^{-a\left(z+z'\right)}\label{eq:C(z,z'_small}
\end{align}
and, from equation (\ref{eq:Isp(L)}),
\begin{align}
    I_\mathrm{sp}\left(L\right) &= \hbar\omega_0\frac{3\pi\Gamma\beta_{k'}}{k^2\alpha_{k'}}\left(n'_0L\right)^2\nonumber \\
    &= 2\hbar\omega_0\beta_{k'}n'_0L\left(aL\right).\label{eq:Isp(L)_small}
\end{align}

Finally, equation (\ref{eq:Isp(L)_small}) can be transformed to a radiation rate $I$ (in units of $\mathrm{s}^{-1}$) through a multiplication with $\lambda L/\left(\hbar\omega_0\right)$ to yield (assuming thermal equilibrium and $\alpha_{k'}=\beta_{k'}$ through $T_2=\infty$)
\begin{align}
    I &= \frac{1}{2}I_0n^2\tanh^2\left(\frac{\hbar\omega_0}{2k_\mathrm{B}T_\mathrm{kin}}\right)\nonumber \\
    &\simeq 2I_0n^2\left(\frac{\hbar\omega_0}{4k_\mathrm{B}T_\mathrm{kin}}\right)^2,\label{eq:I_small}
\end{align}
where $I_0=3\Gamma\lambda/\left(8\pi L\right)$ is the single-mode independent-atom emission rate and $n=2\left(n'_a+n'_b\right)\lambda L^2$ is the total number of atoms. We thus recover, up to a factor of 2, the result first obtained through other means by Dicke for a coherent system initially at thermal equilibrium when $\hbar\omega_0\ll k_\mathrm{B}T_\mathrm{kin}$ (see equations 37, 37a and 44 in \citep{Dicke1954}).   

\section{Collision cross-section}\label{app:cross_section}

We first consider a two-atom system initially in the state
\begin{align}
    \Ket{\psi_0} = \Ket{\mathbf{P},\Delta\mathbf{p}}\Ket{bb}\Ket{0},\label{eq:|P>|bb>}
\end{align}
where $\mathbf{P}=\mathbf{p}_1+\mathbf{p}_2$ and $\Delta\mathbf{p}=\left(\mathbf{p}_2-\mathbf{p}_1\right)/2$ with $\mathbf{p}_1$ and $\mathbf{p}_2$ the linear momenta of the first and second atoms, respectively, and the last state $\Ket{0}$ stands for the vacuum radiation field. We will assume that the two atoms are at resonance, i.e., $\Delta\mathbf{p}=0$ initially. We rewrite the interaction term of the Hamiltonian in equation (\ref{eq:V}) as follows
\begin{align}
    \hat{V} = -i\sum_q\frac{1}{2}\hbar\Omega_q&\sum_{j=1}^2\left[\hat{R}_j^+\hat{T_q}\left(\hbar\mathbf{k}_q\right)\hat{a}_q-\hat{R}_j^-\hat{T_q}^\dagger\left(\hbar\mathbf{k}_q\right)\hat{a}_q^\dagger\right]\label{eq:Vapp}
\end{align}
with $\hat{T_q}\left(\hbar\mathbf{k}_q\right)=e^{i\mathbf{k}_q\cdot\hat{\mathbf{r}}}$ the linear momentum translation operator \cite{Haroche2008}.

Using a first-order perturbation expansion, it can be shown that the state of the system after the emission of a first photon is composed of a superposition of states of the form
\begin{align}
    \Ket{\psi_{+}} &= \frac{1}{\sqrt{2}}\left(\Ket{\hbar\mathbf{k}_r/2}\Ket{ab}+\Ket{-\hbar\mathbf{k}_r/2}\Ket{ba}\right)\Ket{\mathbf{P}-\hbar\mathbf{k}_r}\Ket{1_r}\label{eq:psi_+r}
\end{align}
that differ only in the radiation mode of the emitted photon. In equation (\ref{eq:psi_+r}) the radiation state $\Ket{1_r}$ implies that one photon occupies mode $r$ and all other modes are empty. After a short time the photon leaves the system and $\Ket{1_r}$ is replaced by $\Ket{0}$. 

We can express this state using the position basis $\Ket{\mathbf{R},\Delta\mathbf{r}}$, where $\mathbf{R}=\left(\mathbf{r}_1+\mathbf{r}_2\right)/2$ and $\Delta\mathbf{r}=\mathbf{r}_2-\mathbf{r}_1$ are conjugate to $\mathbf{P}$ and $\Delta\mathbf{p}$, respectively, to get
\begin{align}
    \Ket{\psi_{+}} &\propto \int d^3R\,e^{i\left(\mathbf{P}/\hbar-\mathbf{k}_r\right)\cdot\mathbf{R}}\,\Ket{\mathbf{R}}\int d^3\Delta r\,\Ket{\Delta\mathbf{r}}\nonumber\\
    &\times\frac{1}{\sqrt{2}}\left(e^{i\mathbf{k}_r\cdot\Delta\mathbf{r}/2}\Ket{ab}+e^{-i\mathbf{k}_r\cdot\Delta\mathbf{r}/2}\Ket{ba}\right)\Ket{0}.\label{eq:|psi_+>_position}
\end{align}
We recognize on the last line the internal state $\Ket{1,0}_\theta$ of equation (\ref{eq:|1,0>}) when $\mathbf{k}_r$ makes an angle $\theta$ with $\Delta\mathbf{r}$. The state orthogonal to $\Ket{\psi_{+}}$, and related to $\Ket{0,0}_\theta$ in equation (\ref{eq:|0,0>}), is
\begin{align}
    \Ket{\psi_{-}} &\propto \int d^3R\,e^{i\left(\mathbf{P}/\hbar-\mathbf{k}_r\right)\cdot\mathbf{R}}\,\Ket{\mathbf{R}}\int d^3\Delta r\,\Ket{\Delta\mathbf{r}}\nonumber\\
    &\times\frac{1}{\sqrt{2}}\left(e^{i\mathbf{k}_r\cdot\Delta\mathbf{r}/2}\Ket{ab}-e^{-i\mathbf{k}_r\cdot\Delta\mathbf{r}/2}\Ket{ba}\right)\Ket{0}.\label{eq:|psi_->_position}
\end{align}

As discussed in Sec.~\ref{sec:steady-state}, after the transient regime for systems containing a large number of atoms, states from the dark edge are overwhelmingly populated and every system is essentially in dark states such as the one given in equation (\ref{eq:|psi_->_position}). Let us then consider two such systems $A$ and $B$ entering into a collision. We choose a reference frame at rest with the center-of-mass for the two systems such that $\mathbf{R}_A=-\mathbf{R}_B=-\mathbf{R}_0$. For the 21~cm line of the hydrogen atom we can assume that $\left|\mathbf{P}_A\right|,\,\left|\mathbf{P}_B\right|\gg\hbar k_r$ irrespective of the radiation mode. To lighten the notation we omit the spatial states $\Ket{\mathbf{R}_A}$, $\Ket{\Delta\mathbf{r}_A}$, etc., and the vacuum $\Ket{0}$ state to focus on the internal states, such that we can write for the colliding system   
\begin{align}
    \Ket{\Psi} &= \Ket{\psi_{-}}_A \Ket{\psi_{-}}_B\nonumber\\
    &\propto e^{i\left(\mathbf{P}_B-\mathbf{P}_A\right)\cdot\mathbf{R}_0/\hbar}\left(\Ket{ab}_A-\Ket{ba}_A\right) \left(\Ket{ab}_B-\Ket{ba}_B\right)\nonumber\\
    &\propto e^{i\left(\mathbf{P}_B-\mathbf{P}_A\right)\cdot\mathbf{R}_0/\hbar}\left(\Ket{ab}_A\Ket{ab}_B+\Ket{ba}_A\Ket{ba}_B-\Ket{ab}_A\Ket{ba}_B-\Ket{ba}_A\Ket{ab}_B\right).\label{eq:|Psi>}
\end{align}

For high relative speed between colliding atoms, i.e., when $\left|\mathbf{P}_B-\mathbf{P}_A\right|^2/2m_\mathrm{r}\gg \hbar\omega_0$ ($m_\mathrm{r}$ is the reduced mass), it is expected that the scattering amplitude will be independent of their internal states (see section \ref{app:independence} below). If two atoms enter into a collision, say the first atoms of both pairs, then under the Born approximation, which is well-suited at high energies, each term in equation (\ref{eq:|Psi>}) will bring a scattering amplitude of the type \cite{Cohen-Tannoudji2018b, Taylor2006}
\begin{align}
    f\left(\mathbf{k}_{\mathrm{in}},\mathbf{k}_{\mathrm{out}}\right) = -\frac{m_\mathrm{r}}{2\pi\hbar^2}\int d^3 r^\prime e^{-i\mathbf{k}_{\mathrm{out}}\cdot \mathbf{r}^\prime}V_1\left(\mathbf{r}^\prime\right) e^{i\mathbf{k}_{\mathrm{in}}\cdot \mathbf{r}^\prime}\label{eq:f}
\end{align}
with $\mathbf{k}_{\mathrm{in}}=\left(\mathbf{P}_B-\mathbf{P}_A\right)/\hbar$, $\mathbf{k}_{\mathrm{out}}$ the scattered wave vector and $V_1\left(\mathbf{r}\right)$ the component of the atomic interaction potential independent of the internal states. It therefore follows from equation (\ref{eq:|Psi>}) that under these circumstances the collision cross-section $\sigma\left(\mathbf{k}_{\mathrm{in}},\mathbf{k}_{\mathrm{out}}\right)=0$ since it consists of the square of the sum of the four corresponding scattering amplitudes\footnote{An exchange term should be added to equation (\ref{eq:|Psi>}) to account for the symmetrization process stemming from the indistinguishability of the two identical atoms during the collision \cite{Cohen-Tannoudji2018b, Taylor2006}. For simplicity, we only consider the direct term since the same result (i.e., the vanishing of the corresponding scattering amplitude) is obtained for the exchange term.}.

Finally, we note that the cancellation of the collision cross-section is independent of temperature once entanglement is established. The entanglement between atoms can only be made more difficult through an increase in temperature since greater thermal speeds will reduce the likelihood that atoms can interact with a common radiation field. Lower temperatures can only help in the process.


\subsection{Generalization to arbitrarily large atomic ensembles}\label{sec:N_ensembles}

The vanishing of the scattering amplitude for two two-atom ensembles rests on the fact that sum of the probability amplitudes in the overall state of the colliding system equals zero (i.e., in equation \ref{eq:|Psi>}). It is straightforward to verify that this will happen whenever at least one of the systems is in the anti-symmetric state given in equation (\ref{eq:|psi_->_position}). That is, we also get a vanishing scattering cross-section for collisions involving, say, any of the symmetric states $\ket{bb}_A, \left(e^{i\mathbf{k}_r\cdot\Delta\mathbf{r}_A/2}\ket{ab}_A+e^{-i\mathbf{k}_r\cdot\Delta\mathbf{r}_A/2}\ket{ba}_A\right)/\sqrt{2}$ and $\ket{aa}_A$ with $\Ket{\psi_{-}}_B$ of equation (\ref{eq:|psi_->_position}). We now show that this result can be generalized to arbitrarily large entangled atomic ensembles.

First, we note that for a system of $N$ atoms any Dicke state that is not totally-symmetric has the sum of its probability amplitudes equal to zero. To verify this we write the totally-symmetric state for a given quantum number $m$ as
\begin{align}
    \Ket{r=N/2,m} = \sqrt{\frac{\left(N/2+m\right)!\left(N/2-m\right)!}{N!}}\sum_j \Ket{j},\label{eq:|N/2,m>}
\end{align}
where the summation is on all possible atomic state $\Ket{j}$ having $N/2-m$ atoms in the lower $\Ket{a}$ and $N/2+m$ in the upper $\Ket{b}$ states. Similarly, the non-totally-symmetric $\Ket{r<N/2,m}$ states are also expressed as linear superpositions of the type
\begin{align}
    \Ket{r<N/2,m} = \sum_j d_j\Ket{j}.\label{eq:|r<N/2,m>}
\end{align}
However, since the states on the left-hand sides of equations (\ref{eq:|N/2,m>}) and (\ref{eq:|r<N/2,m>}) are orthogonal we can write
\begin{align}
    \sqrt{\frac{N!}{\left(N/2+m\right)!\left(N/2-m\right)!}}\braket{r=N/2,m|r<N/2,m} &= \sum_{j,k}d_k\braket{j|k}\nonumber\\
    &= \sum_j d_j\nonumber\\
    &= 0,\label{eq:<N/2,n|r<N/2,m>}
\end{align}
which establishes our previous assertion that any Dicke state that is not totally-symmetric has the sum of its probability amplitudes equal to zero. 

This statement also holds for the products of any two Dicke states of arbitrary quantum numbers, as long as one of the states is not totally-symmetric, since
\begin{align}
    \Ket{r^\prime,m^\prime}\Ket{r,m} &= d^\prime_1\Ket{1^\prime}\left(\sum_j d_j\Ket{j}\right)+d^\prime_2\Ket{2^\prime}\left(\sum_j d_j\Ket{j}\right)+\dots\label{eq:|r'm'>|rm>}
\end{align}
yields the sum of probability amplitudes
\begin{align}
    d^\prime_1\left(\sum_jd_j\right)+d^\prime_2\left(\sum_j d_j\right)+\dots = 0\label{eq:d'sum_d}
\end{align}
from equation (\ref{eq:<N/2,n|r<N/2,m>}) with $\Ket{r,m}$ a non-totally symmetric state. Our generalization of the vanishing cross-section between two entangled atomic ensembles, with one in a non-totally-symmetric state, thus follows from this realization.

We also note that the product of two totally symmetric states yields
\begin{align}
    \Ket{N^\prime/2,m^\prime} & \Ket{N/2,m} = \nonumber\\ 
    & \sqrt{\frac{\left(N^\prime/2+m^\prime\right)!\left(N^\prime/2-m^\prime\right)!}{N^\prime!}}\sqrt{\frac{\left(N/2+m\right)!\left(N/2-m\right)!}{N!}}\sum_{j^\prime j} \Ket{j^\prime}\Ket{j}\label{eq:one_symm}
\end{align}
with the corresponding sum of the probability amplitudes
\begin{align}
    &\sqrt{\frac{\left(N^\prime/2+m^\prime\right)!\left(N^\prime/2-m^\prime\right)!}{N^\prime!}} \sqrt{\frac{\left(N/2+m\right)!\left(N/2-m\right)!}{N!}}\sum_{j^\prime j} = \nonumber\\ 
    & \qquad\qquad\qquad\qquad\sqrt{\frac{N^\prime!}{\left(N^\prime/2+m^\prime\right)!\left(N^\prime/2-m^\prime\right)!}}\sqrt{\frac{N!}{\left(N/2+m\right)!\left(N/2-m\right)!}}.\label{eq:factor_symm}
\end{align}

To make these results more explicit, we return to the problem of colliding entangled ensembles and consider systems of $N_A$ and $N_B$ atoms but limit ourselves to the $N_\alpha$ states for $m_\alpha=-N_\alpha/2+1$ ($\alpha=A,B$) when only one atom is excited. It can be shown that the following are appropriate internal dark states (i.e., $r_\alpha=N_\alpha/2-1$)
\begin{align}
    \Ket{N_\alpha/2-1,-N_\alpha/2+1,q} = \frac{1}{\sqrt{q\left(q+1\right)}}\left(\sum_{j=1}^q\Ket{b_j}_\alpha-q\Ket{b_{q+1}}_\alpha\right),\label{eq:N_dark-states}
\end{align}
where $1\le q \le N_\alpha-1$ is used to label the $N_\alpha-1$ dark states, $\Ket{b_j}$ implies that only the $j^\mathrm{th}$ atom is in the upper state (all others are in the lower $\Ket{a}$ state) and we once again assume that $\left|\mathbf{P}_\alpha\right|\gg\hbar k_r$. Clearly, these states have the sum of their probability amplitudes equal to zero since $q-\sum_{j=1}^q=0$. The symmetric states are
\begin{align}
    \Ket{N_\alpha/2,-N_\alpha/2+1} = \frac{1}{\sqrt{N_\alpha}}\sum_{j=1}^{N_\alpha}\Ket{b_j}_\alpha,\label{eq:N_symmetric}
\end{align}
which have their sum of probability amplitudes equal to $\sqrt{N_\alpha}$.

The overall state for the scattering between two non-totally-symmetric entangled states labeled $q$ (for system $A$) and $q^\prime$ (for system $B$) is therefore 
\begin{align}
    \Ket{\Psi_N} & \propto  e^{i\left(\mathbf{P}_B-\mathbf{P}_A\right)\cdot\mathbf{R}_0/\hbar}\left( qq^\prime\Ket{b_{q+1}}_A\Ket{b_{q^\prime+1}}_B+\sum_{j,j^\prime=1}^{q,q^\prime}\Ket{b_j}_A\Ket{b_{j^\prime}}_B\right.\nonumber\\
    & \qquad\qquad\qquad\qquad\left.-q\sum_{j^\prime=1}^{q^\prime}\Ket{b_{q+1}}_A\Ket{b_{j^\prime}}_B-q^\prime\sum_{j=1}^q\Ket{b_j}_A\Ket{b_{q^\prime+1}}_B \right).\label{eq:Psi_N} 
\end{align}
We find that $\Ket{\Psi_N}$ has exactly the same structure as $\Ket{\Psi}$ in equation (\ref{eq:|Psi>}) for the two-atom ensembles, but with the difference that each term in the parentheses has a weight of $qq^\prime$. However, the outcome is the same as the scattering amplitude vanishes at high energies because of its independence on the nature of the internal states. This result will hold in the more general case where the entangled systems are in a superposition of dark states (i.e., including more than one value for $q$ in equation \ref{eq:N_dark-states} for a given system) or, as mentioned above, one of the colliding systems is in a totally-symmetric state.

Finally, when both systems are in totally-symmetric states we find 
\begin{align}
    \Ket{\Psi_N} & \propto  e^{i\left(\mathbf{P}_B-\mathbf{P}_A\right)\cdot\mathbf{R}_0/\hbar}\sqrt{\frac{1}{N_AN_B}}\sum_{j,j^\prime}\Ket{b_j}_A\Ket{b_{j^\prime}}_B\label{eq:Psi_N-symm}
\end{align}
with a corresponding scattering cross-section $N_AN_B\sigma_0$, where $\sigma_0$ is the cross-section for two independent atoms.

We thus conclude that at high relative speed two entangled systems will pass through each other, unimpeded, during a collision whenever at least one is in a non-totally-symmetric state. This is due to the destructive interference resulting from the entanglement within that system. In contrast, when both systems are in totally-symmetric states the ``normal'' scattering cross-section $\sigma_0$ for the collision between two independent atoms is amplified by a factor equal to the product of the numbers of atoms in the two entangled ensembles (for the last case considered above; for the more general case, the amplification is given by the square of the factor found on the right-hand side of equation \ref{eq:factor_symm}). It also follows that these outcomes would hold for any other types of high-energy scattering processes (e.g., involving cosmic rays, etc.). For example, atoms in darkened entangled systems cannot become ionized through collisions even in environments where the gas temperature would be high enough to do so for an independent atom.

\subsection{Independence on internal states for high energy scattering}\label{app:independence}

In the center-of-mass of the system, the Hamiltonian for the scattering of two hydrogen atoms in the electronic ground state can be expressed as  \cite{Zygelman2003}
\begin{align}
    \hat{H} = \frac{\hat{p}^2}{2m_\mathrm{r}}+\hat{V}_1\left(\mathbf{r}\right)+\hat{H}_\mathrm{hf} \label{eq:Hcoll}
\end{align}
with $\mathbf{r}$ and $\mathbf{p}$ the relative position and (half of) the relative linear momentum between the two atoms, respectively. The hyperfine Hamiltonian results from the intra-atomic interactions due to the Fermi contact term \cite{Cohen-Tannoudji2018b}, which dominates over any other interatomic spin coupling term, and is thus given by
\begin{align}
    \hat{H}_\mathrm{hf} = \mathcal{A}\left(\hat{\mathbf{I}}_1\cdot\hat{\mathbf{S}}_1+\hat{\mathbf{I}}_2\cdot\hat{\mathbf{S}}_2\right),\label{eq:Hhf}
\end{align}
where $\hat{\mathbf{I}}_j$ and $\hat{\mathbf{S}}_j$ are the proton and electron spins for atom $j$, and $\mathcal{A}=\omega_0/\hbar$ a constant. In equation (\ref{eq:Hcoll}) (and \ref{eq:f}), the interatomic potential $\hat{V}_1\left(\mathbf{r}\right)$ is independent of the internal states (and spins). 

From a classical standpoint, the turning point $r_\mathrm{min}$ for a collision at energy $E_\mathrm{kin}=p^2/2m_\mathrm{r}$ will be approximately determined by $V_1\left(r_\mathrm{min}\right)\sim E_\mathrm{kin}$ \cite{Purcell1956}. In cases such as the bullet cluster \cite{Clowe2006,Cirelli2024}, with collision speeds $\sim10^3\,\mathrm{km\,s^{-1}}$, we find that $E_\mathrm{kin}\sim 10^9\hbar\omega_0$ and therefore $V_1\left(r_\mathrm{min}\right) \gg \hbar\omega_0$, or equivalently $V_1\left(r_\mathrm{min}\right)\gg H_\mathrm{hf}$ for the 21~cm atomic hydrogen line. It follows that the hyperfine Hamiltonian is completely negligible in the region where the scattering occurs. 

Since the internal states are determined by the atomic spins, we observe a decoupling between the internal and external degrees of freedom on account of the large separation in the corresponding energy scales. In other words, the scattering amplitude is largely independent of the atoms' internal state and can be evaluated with equation (\ref{eq:f}). 




\end{appendices}

\end{document}